\documentclass{llncs}

\usepackage{times,subfigure}
\usepackage{amsmath}
\usepackage{amsfonts}
\usepackage{amssymb}
\usepackage{graphicx}
\usepackage{color}
\usepackage{algorithm}
\usepackage{algorithmic}
\usepackage{url}
\newcommand{\trimmer}{\textsc{CGHtrimmer}}
\newcommand{\dnacopy}{\textsc{CBS}}
\newcommand{\cghseg}{\textsc{CGHseg}}
\newcommand{\field}[1]{\mathbb{#1}}

\begin{document}

\title{ CGHTRIMMER: Discretizing noisy Array CGH Data }

\titlerunning{ Algorithms for Denoising aCGH Data}  % abbreviated title (for running head)
\author{Charalampos E. Tsourakakis\inst{1},  David Tolliver\inst{1}, Maria A. Tsiarli\inst{2}, Stanley Shackney\inst{3}, Russell Schwartz\inst{1,4}}
\authorrunning{Tsourakakis et al.}   % abbreviated author list (for running head)

\institute{ School of Computer Science, Carnegie Mellon University, USA\\
\email{ctsourak@cs.cmu.edu}, \email{tolliver@cs.cmu.edu}\\ 
\and
Center for Neuroscience University of Pittsburgh, Center for the Neural Basis of Cognition, USA \\
\email{mat90@pitt.edu} \\
\and
Departments of Oncology and Human Genetics, Drexel University, USA \\
\and
Department of Biological Sciences, Carnegie Mellon University, USA\\
\email{russells@andrew.cmu.edu} }

\maketitle

\begin{abstract}
The development of cancer is largely driven by the gain or loss of
subsets of the genome, promoting uncontrolled growth or disabling
defenses against it.  Identifying genomic regions whose DNA copy
number deviates from the normal is therefore central to understanding
cancer evolution. Array-based comparative genomic hybridization (aCGH)
is a high-throughput technique for identifying DNA gain or loss by
quantifying total amounts of DNA matching defined probes relative to
healthy diploid control samples.  Due to the high level of noise in
microarray data, however, interpretation of aCGH output is a difficult
and error-prone task.

In this work, we tackle the computational task of inferring the DNA
copy number per genomic position from noisy aCGH data.  We
propose \trimmer, a novel segmentation method that uses a fast dynamic
programming algorithm to solve for a least-squares objective function
for copy number assignment.  \trimmer\ consistently achieves superior
precision and recall to leading competitors on benchmarks of synthetic
data and real data from the Coriell cell lines.  In addition, it finds
several novel markers not recorded in the benchmarks but plausibly
supported in the oncology literature.  Furthermore, \trimmer\ achieves
superior results with run-times from 1 to 3 orders of magnitude faster
than its state-of-art competitors.

 \trimmer\ provides a new alternative for the problem of aCGH discretization that provides superior detection of fine-scale regions of gain or loss yet is fast enough to process very large data sets in seconds. It thus meets an important need for methods capable of handling the vast amounts of data being accumulated in high-throughput studies of tumor genetics.  

\end{abstract}

\section{Introduction}
\label{sec:intro}
Tumorigenesis is a complex phenomenon often characterized by the
successive acquisition of combinations of genetic aberrations that
result in malfunction or disregulation of genes.  There are many forms
of chromosome aberration that can contribute to cancer development,
including polyploidy, aneuploidy, interstitial deletion, reciprocal
translocation, non-reciprocal translocation, as well as amplification,
again with several different types of the latter (e.g., double minutes, HSR
and distributed insertions \cite{albertson}).  Identifying the specific
recurring aberrations, or sequences of aberrations, that characterize
particular cancers provides important clues about the genetic basis of
tumor development and possible targets for diagnostics or
therapeutics.  Many other genetic diseases are also characterized by
gain or loss of genetic regions, such as Down Syndrome (trisomy
21)~\cite{downsyndrome}, Cri du Chat (5p deletion)~\cite{criduchat}, and
Prader-Willi syndrome (deletion of 15q11-13)~\cite{praderwilli} and
recent evidence has begun to suggest that inherited copy number
variations are far more common and more important to human health than
had been suspected just a few years ago~\cite{CNVs}.  These facts have
created a need for methods for assessing DNA copy number variations
in individual organisms or tissues.

\begin{figure} 
\centering
\includegraphics[width=.6\textwidth]{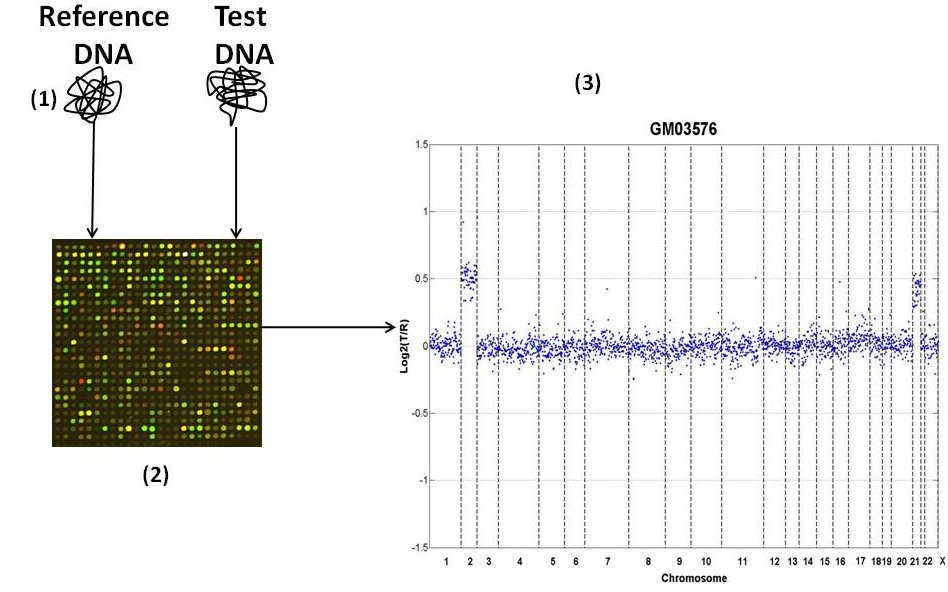}
\caption{Schematic representation of array CGH.  Genomic DNA from two cell populations (1)
is differentially labeled and hybridized in a microarray (2). Typically the reference 
DNA comes from a normal subject. For humans  this means that the reference DNA comes
from a normal diploid genome. The ratios on each spot 
are measured and normalised so that the median $\log_2$ ratio is zero. 
The final result is an ordered tuple  containing values 
of the fluorescent ratios  in each genomic position per each chromosome. This is shown in (3)
where we see the genomic profile of the cell line GM05296 \cite{coriell}. 
The problem of denoising array CGH  data is to infer the true DNA copy number $T$ 
per genomic position from a set of noisy measurements of the quantity $\log_2{ \frac{T}{R}}$,
where $R$=2 for normal diploid humans. }
\label{fig:arraycgh}
\end{figure}

In this work, we focus specifically on array-based comparative genomic
hybridization
(aCGH) \cite{bignell,pollack,Kallioniemi:1992,Genomic04high}, a method
for copy number assessment using DNA microarrays that remains, for the
moment, the leading approach for high-throughput typing of copy number
abnormalities.  The technique of aCGH is schematically represented in
Figure~\ref{fig:arraycgh}.  A test and a reference DNA sample are
differentially labeled and hybridized to a microarray and the ratios
of their fluorescence intensities is measured for each spot. A typical
output of this process is shown in Figure~\ref{fig:arraycgh} (3),
where the genomic profile of the cell line GM05296 \cite{coriell} is
shown for each chromosome. The x-axis corresponds to genomic position
and the y-axis corresponds to a noisy measurement of the ratio
$\log_2{ \frac{T}{R}}$ for each genomic position.  For healthy diploid
organisms, $R$=2 and $T$ is the DNA copy number we want to infer from
the noisy measurements.
For more details on the use of aCGH to detect different types of 
chromosomal aberrations, see \cite{albertson}.

Converting raw aCGH log fluorescence ratios into discrete DNA copies
numbers is an important but non-trivial problem in using aCGH to
study cancer progression.  Finding DNA regions that consistently
exhibit chromosomal losses or gains in cancers provides a crucial
means for locating the specific genes involved in development of
different cancer types.  It is therefore important to distinguish,
when a probe shows unusually high or low fluorescence, whether that
aberrant signal reflects experimental noise or a probe that is truly
found in a segment of DNA that is gained or lost.  Furthermore,
successful discretization of array CGH data is crucial for
understanding the process of cancer evolution, since discrete inputs
are required for a large family of successful evolution algorithms,
e.g., \cite{DBLP:journals/jcb/DesperJKMPS99,DBLP:journals/jcb/DesperJKMPS00}.
It is worth noting that manual annotation of such regions, even if
possible \cite{coriell}, is tedious and prone to mistakes due to
several sources of noise (impurity of test sample, noise from array
CGH method, etc.).

Many algorithms and objective functions have thus been proposed for
the problem of discretizing and segmenting aCGH data.  Many methods,
starting with Fridlyand et al.~\cite{1016476}, treat aCGH segmentation
as a hidden Markov model (HMM) inference problem.  The HMM approach
has since been extended in various ways, such as through the use of
Bayesian HMMs \cite{guha}, incorporation of prior knowledge of
locations of DNA copy number polymorphisms~\cite{citeulike:789789}, and
the use of Kalman filters~\cite{DBLP:conf/recomb/ShiGWX07}.  Other
approaches include wavelet decompositions~\cite{hsu}, quantile
regression~\cite{citeulike:774308}, expectation-maximization in
combination with edge-filtering~\cite{citeulike:774287}, genetic
algorithms~\cite{DBLP:journals/bioinformatics/JongMMVY04},
clustering-based methods~\cite{xing,citeulike:773210}, variants on
Lasso regression~\cite{citeulike:2744846,huang}, and various
problem-specific Bayesian~\cite{barry}, likelihood~\cite{1093217}, and
other statisical models~\cite{DBLP:conf/recomb/LipsonABLY05}.  A
dynamic programming approach, in combination with expectation
maximimization, has been previously used by Picard et
al.~\cite{picard2}.  Lai et al.~\cite{1181383} and Willenbrock et
al.~\cite{citeulike:387317} have conducted extensive experimental
analysis of the range of available methods, with two in particular
standing out as the leading approaches in practice.  One of these top
methods is \cghseg\ \cite{picard}, which assumes that a given CGH
profile is a Gaussian process whose distribution parameters are
affected by abrupt changes at unknown coordinates/breakpoints.  The
other is Circular Binary Segmentation~\cite{olshen} (CBS), a
modification of binary segmentation, originally proposed by Sen and
Srivastava~\cite{sensrivastava}, which uses a statistical comparison
of mean expressions of adjacent windows of nearby probes to identify
possible breakpoints between segments combined with a greedy algorithm
to locally optimize breakpoint positions.

Our main contribution in this work is a new algorithm, \trimmer, for
denoising and segmentation of aCGH data.  We develop a novel objective
function for the problem based on least-squares minimization of errors
combined with a regularization parameter to favor contiguity across
segments.  We show how to solve efficiently for this objective
function through dynamic programming.  We then validate the method, in
comparison to the leading CBS and \cghseg\ methods, on a combination of
synthetic and real benchmarks.  Finally, we show that \trimmer\ yields
superior accuracy in identifying known breakpoints while performing
one to three orders of magnitude faster than the comparative methods.
The remainder of the paper is organized as follows:
Section~\ref{sec:prop} presents our proposed method and its
theoretical analysis. Section~\ref{sec:expe} describes the
experimental setup and shows the experimental results of our method
compared to two state-of-art methods \cite{olshen,picard} on both
synthetic and real aCGH data. Section~\ref{sec:conc} concludes the
paper with a brief summary and discussion.

% Figure~\ref{fig:cover} shows the speedup obtained using \trimmer\ versus
%the leading CBS and \cghseg\ methods. Specifically, Figure~\ref{fig:cover} plots 
%the speedup versus a given pair of chromosome and cell line of the Coriell data,
%a dataset considered to be the ``gold'' standard \cite{coriell}. 
%The red and blue points show the speedup when we run  \trimmer\ algorithm 
%versus the Circular Binary Segmentation (CBS) \cite{olshen} and \cghseg\ \cite{picard}
%respectively. As we observe \trimmer\ is from 1 to 3 orders of magnitude faster compared to those methods. 
%The quality of the results is consistently at least as good as the results of the competitors
%as shown in Section~\ref{sec:expe}.

\section{Proposed Method}
\label{sec:prop}
\begin{table}[!t]
\centering
\begin{tabular}{c|c}
 Symbol       & Description  \\ \hline
$p_i$    &  measurement of $\log_2{\frac{T}{R}}$ in the $i$-th probe       \\
     n      & number of probes \\
     T      &  DNA copy number   \\ 
     R      & reference value, equal to 2 for diploid organisms   \\
     K      & number of segments to be fitted \\ 
  $\lambda$ & regularization parameter \\ 
$\mu_{(j)}$ & Average of points $\{ p_1,\ldots, p_j\}$, $\mu_{(j)} = \sum_{i=1}^j p_i/j$ \\
$S_{(j)}$   & Sum of squared errors $\sum_{i=1}^j ( p_i - \mu_{(j)})^2 $ \\ \hline
CBS    &  Circular binary segmentation \cite{olshen} \\ 
\cghseg & Picard et al. \cite{picard} \\ \hline
\end{tabular}
\caption{Symbols and abbreviations used.}
{\label{tab:symbols}}
\end{table}

We formulate the problem of denoising aCGH data as a least squares problem with a penalty/regularization
term for the number of segments. Table~\ref{tab:symbols} shows the symbols used throughout the paper for convenience.
Let $P=\{p_1,\ldots,p_n\}$ be a set of $n$ points, $p_i \in \field{R}$. Our goal is to find a sequence of piecewise 
constant segments which minimize the sum of squared errors per segment 
and the number of segments $K$.
Ideally, our algorithm will fit the data with piecewise constant
segments corresponding to the true DNA copy number at each genomic position. 
The intuition for penalizing the number of segments is that we expect a 
strong spatial coherence between nearby probes.  Gains and losses are likely to occur for DNA segments covering many probes, and adjacent probes should therefore usually have the same copy number. 
We avoid making explicit assumptions about the statistical nature of the signal (although we note that the
least squares method can be interpreted as a maximization of the likelihood function under
the assumption that noise derives from identically distributed Gaussian random variables) since we do not believe we have a strong empirical basis for exactly capturing the true correlation structure of aCGH data. 
To solve this optimization problem, we define the key quantity $OPT_i$, given by the following equation:

\begin{equation} 
OPT_i = \left\{ 
\begin{array}{l l}
  0 & \quad \mbox{\text{if} $i=0$}\\
  \min_{1 \leq j \leq i} OPT_{j-1}+\sum_{k=j}^i \left(p_k - \frac{\sum_{m=j}^i p_m}{i-j+1} \right)^2+ \lambda & \quad \mbox{\text{if} $i>0$}\\ \end{array} \right.   
  \label{eq:eqrec}
\end{equation}

The recursion equation~\ref{eq:eqrec} has a straightforward interpretation:
$OPT_i$ is equal to the minimum cost of fitting a set of piecewise constant segments 
from point $p_1$ to $p_i$ given that the last change in copy numbers occured between points $p_{j-1}$ and $p_j$ plus the cost of fitting a segment is $\lambda$. 
The second term is the minimum squared error for fitting a constant segment 
on points $\{p_j,\ldots,p_i\}$, which is obtained for the constant segment with value 
equal to the average intensity of the points in the segment, i.e., $\frac{\sum_{m=j}^i p_m}{i-j+1}$.
Recursion~\ref{eq:eqrec} directly implies a dynamic programming algorithm, the \trimmer\ algorithm presented as Algorithm 1. For each point $p_i$, we find a point $p_j$ such that $j$ is the minimum
index over all points before $i$ such that points $p_j$  through $p_i$ belong 
to the same segment.
Since aCGH data are given in the log scale, we first exponentiate the points, 
then fit the constant segment by taking the average of the exponentiated values from the hypothesized segment,
and then return to the log domain by taking the logarithm of that constant value. 
Observe that one can fit a constant segment by averaging the log values using Jensen's inequality, but we favor an approach more consistent with the prior work, which typically models the data assuming i.i.d.~Gaussian noise
in the linear domain. 
The algorithm decides which points will be assigned to the same segment by tracing back, starting from the last point $p_n$,
using the breakpoint variables until it assigns the first point $p_1$ to a segment. 
The main computational bottleneck of \trimmer\ is the computation of an auxiliary matrix $M$, an upper diagonal matrix for which $m_{ij}$ is the minimum squared
error of fitting a segment from points $\{p_i,\ldots,p_j\}$.
To avoid a na\"ive algorithm that would simply find the average of those points
and then compute the squared error, resulting in $O(n^3)$ time,
we take advantage of the following theorem:

\begin{algorithm}
\begin{algorithmic}
\REQUIRE Points $P= \{ p_1,\ldots,p_n \}$
\REQUIRE Regularization parameter $\lambda$ 
\STATE   Compute an $n \times n$ matrix $M$, where 
\STATE   $M_{ji} \leftarrow \sum_{k=j}^i \left(p_k - \frac{\sum_{m=j}^i p_m}{i-j+1} \right)^2 $.
\FOR{  $i=1$ to $n$     } 
\STATE  $ \text{OPT}_i  \leftarrow \min_{1 \leq j \leq i}{ OPT_{j-1}+M_{j,i}+ \lambda   }$
\STATE  $ \text{BREAK}_i  \leftarrow arg\min_{1 \leq j \leq i}{ OPT_{j-1}+M_{j,i} + \lambda}$
\ENDFOR 
\STATE tmp $\leftarrow n$
\COMMENT{Assign points to segments}
\WHILE{tmp $\neq 0$}
\STATE Assign points $\{ p_{BREAK_{\text{tmp}}}, \ldots, p_{\text{tmp}} \}$ to one segment
\STATE tmp $\leftarrow BREAK_{\text{tmp}}-1$ 
\ENDWHILE
\end{algorithmic}
\caption{ CHGtrimmer algorithm.}
\end{algorithm}

\begin{algorithm}
\begin{algorithmic}
\REQUIRE Points $P= \{ p_1,\ldots,p_n \}$
\STATE Initialize matrix $A \in \field{R}^{n \times n}$, $A_{ij}=0, i \neq j$
and $A_{ii}= p_i$.
\STATE Initialize matrix $M \in \field{R}^{n \times n}$ with zeros.
%\COMMENT{ A contains the means }
\FOR{ $i=1$ to $n$ }
\FOR{ $j=i+1$ to $n$ }
\STATE $A_{i,j} \leftarrow \frac{j-i}{j-i+1} A_{i,j-1}+ \frac{1}{j-i+1} p_j$
\ENDFOR
\ENDFOR

\FOR{ $i=1$ to $n$ }
\FOR{ $j=i+1$ to $n$ }
\STATE $M_{i,j} \leftarrow M_{i,j-1} + \frac{j-i}{j-i+1}(p_j-A_{i,j-1})^2$ 
\ENDFOR
\ENDFOR

\end{algorithmic}
\caption{ Computing matrix $M$ efficiently in $O(n^2)$ }
\label{alg:alg2}
\end{algorithm}

\begin{theorem}

Let $m_{(j)}$ and $S_{(j)}$ be the average and the minimum squared error of fitting a constant segment 
for points $\{ p_1,\ldots,p_j \}$. Then the following equations hold: 

\begin{equation}
m_{(j)}= \frac{j-1}{j} m_{(j-1)} + \frac{1}{j} p_j
\label{eq:onlinemean}
\end{equation}

\begin{equation}
S_{(j)}= S_{(j-1)} + \frac{j-1}{j} (p_j-m_{(j-1)})^2
\label{eq:onlinevar}
\end{equation}

\end{theorem} 

For a proof, see \cite{knuth}. Equations~\ref{eq:onlinemean} and~\ref{eq:onlinevar} provide us a way to compute means 
and least squared errors online, leading to the $O(n^2)$ Algorithm 2 for computing matrix $M$.

The resulting method has $O(n^2)$ time and space complexity.  The
algorithm needs to store two $n \times n$ matrices, $M$ and $A$, in
addition to the size-$n$ $OPT$ and $BREAK$ matrices needed for the
dynamic programming.  Therefore the space complexity is $O(n^2)$. The
$A$ and $M$ matrices each have $O(n^2)$ entries and require $O(1)$
time to compute each entry, while computing the $OPT_i$ and $BREAK_i$
entries requires at most $O(n)$ time for each of $n$ entries.
Therefore the total running time is $O(n^2)$.

\section{Results}
\label{sec:expe}
\begin{table*}[!htb]
\centering
\begin{tabular}{c|c|c}\hline
        ~                 &      Dataset           &  Availability             \\\hline
\textcolor{red}{$\odot$} &  Lai et al.        & \cite{1181383}                                                           \\ 
      ~                   &                    & \url{http://compbio.med.harvard.edu/}   \\ \hline
\textcolor{red}{$\odot$} & Willenbrock et al. & \cite{citeulike:387317}                                                    \\
    ~                   &                    &  \url{http://www.cbs.dtu.dk/~hanni/aCGH/}                               \\ \hline
\textcolor{cyan}{$\blacksquare$}&  Coriell Cell lines     & \cite{coriell} \\
    ~       &           &  \url{http://www.nature.com/ng/journal/v29/n3/} \\ \hline
\textcolor{cyan}{$\blacksquare$} & Berkeley Breast Cancer & \cite{berkeley} \\             &           & \url{http://icbp.lbl.gov/breastcancer/}  \\ \hline
\end{tabular}
\caption{Datasets, papers and the URLs where the datasets can be downloaded. 
\textcolor{red}{$\odot$} and \textcolor{cyan}{$\blacksquare$}   denote
which datasets are synthetic and real respectively.}
\label{tab:descriptiondataset}
\end{table*}

This section is organized as follows: first we describe the experimental setup and how we trained our model.
Then we present our experimental results on both synthetic and real data.

\subsection{Experimental Setup and Datasets}
\trimmer\ is implemented in MATLAB.  The experiments run in a 4GB RAM,
2.4GHz Intel(R) Core(TM)2 Duo CPU, Windows Vista machine.  Our methods
were compared to existing MATLAB implementations of the CBS algorithm,
available via the Bioinformatics toolbox, and the \cghseg\ algorithm,
generously provided by by Franc Picard \cite{picard}.  \cghseg\ was run
using heteroscedastic model under the Lavielle criterion
\cite{DBLP:journals/sigpro/Lavielle05}.  Additional tests using the
homoscedastic model showed substantially worse performance and are
omitted here.  All methods were compared using previously developed
benchmark datasets, shown in Table~\ref{tab:descriptiondataset}.
Follow-up analysis of detected regions was conducted by manually
searching for significant genes in the Genes-to-Systems Breast Cancer
Database \url{http://www.itb.cnr.it/breastcancer}~\cite{G2SBCD} and
validating their positions with the UCSC Genome Browser
\url{http://genome.ucsc.edu/}. The Atlas of Genetics and
Cytogenetics in Oncology and Haematology
\url{http://atlasgeneticsoncology.org/} was also used to validate the
significance of reported cancer-associated genes.

\begin{figure}
\centering
\includegraphics[width=0.7\textwidth]{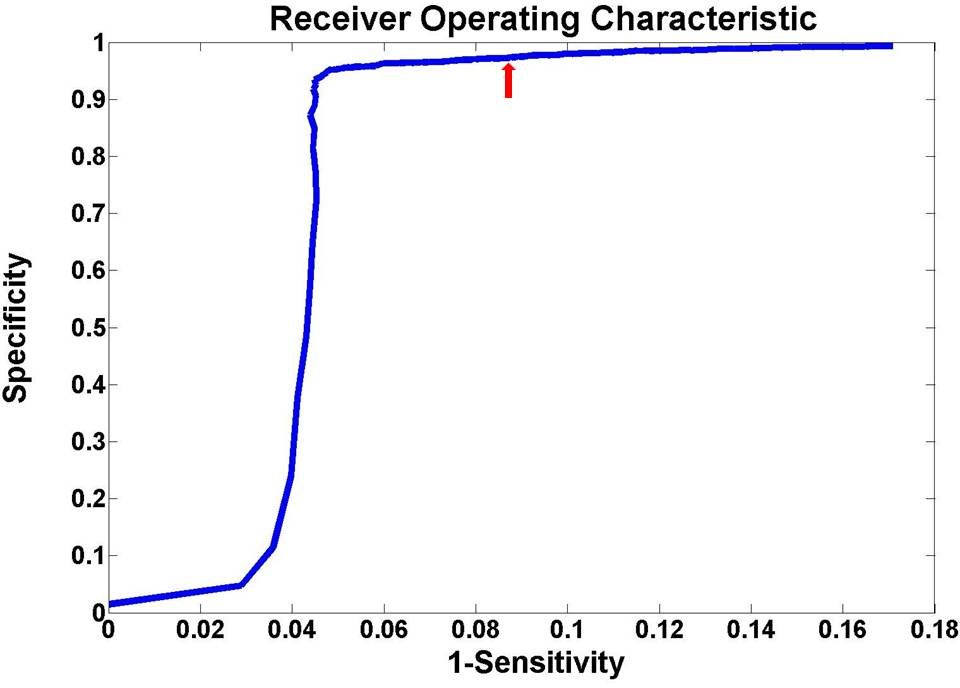}
\caption{ROC curve of \trimmer\ as a function of $\lambda$ on data from \cite{citeulike:387317}. The red arrow indicates
the point (0.91 and 0.98 recall and precision respectively) 
corresponding to $\lambda$=0.2, the value used in all subsequent results.}
\label{fig:roc}
\end{figure}

\subsection{Picking $\lambda$}

The performance of our algorithm depends on the value of the parameter $\lambda$, 
which determines how much each segment ``costs.'' Clearly, there is a tradeoff between
bigger and smaller values: excessively large $\lambda$ will lead the algorithm to output a single segment while excessively small $\lambda$ will result in each point being fit as its own segment.  
We pick our parameter $\lambda$ using data published in \cite{citeulike:387317}.
These data have been generated by modeling real CGH data, thus capturing their 
nature better than other simplified synthetic data and also making them a good training dataset
for our model.  We used this dataset to generate a Receiver Operating Characteristic (ROC) curve 
using values for $\lambda$ ranging from 0 to 4 with increment 0.01 using one of the four datasets in \cite{citeulike:387317} (``above 20''). 
The resulting curve is shown in Figure~\ref{fig:roc}.
We then selected $\lambda=0.2$, which achieves high precision (0.98) and high recall (0.91).
All subsequent results reported were obtained by setting $\lambda$ equal to 0.2.

\begin{figure*}[htp]
  \begin{center}
    \subfigure[ \trimmer ]{\includegraphics[width=0.45\textwidth]{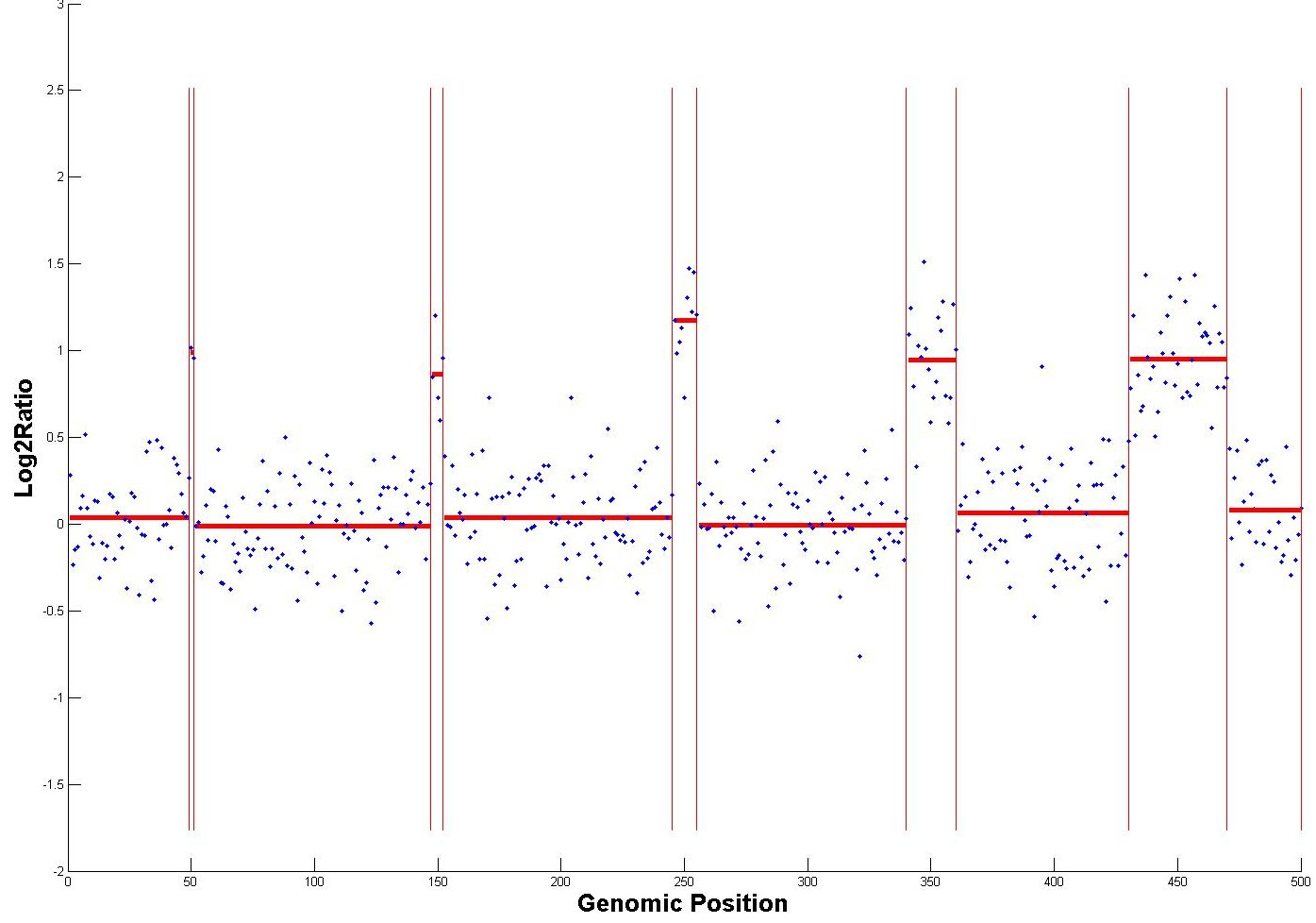}}
    \subfigure[ \dnacopy]{\includegraphics[width=0.45\textwidth]{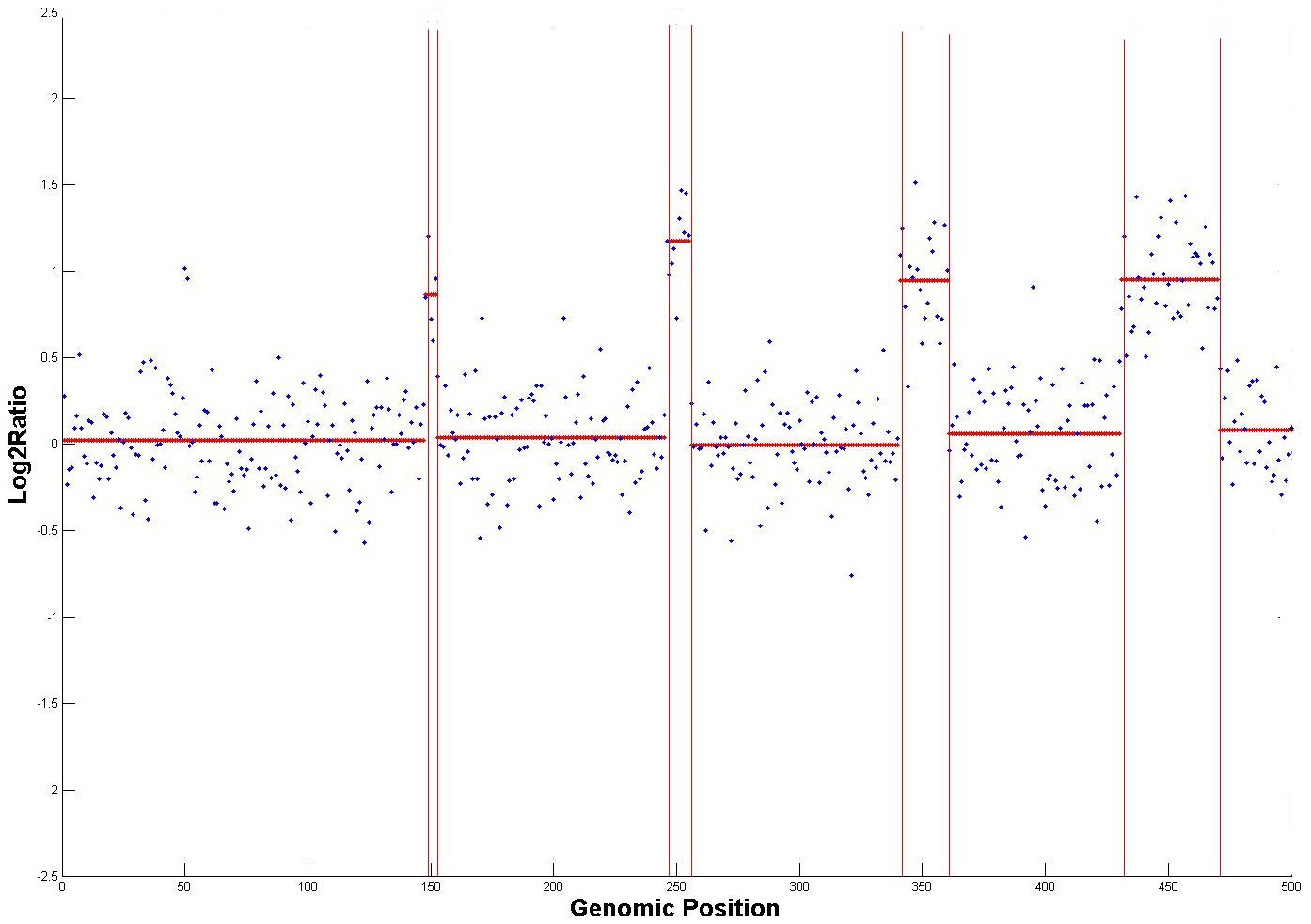}} \\
    \subfigure[ \cghseg]{\includegraphics[width=0.45\textwidth]{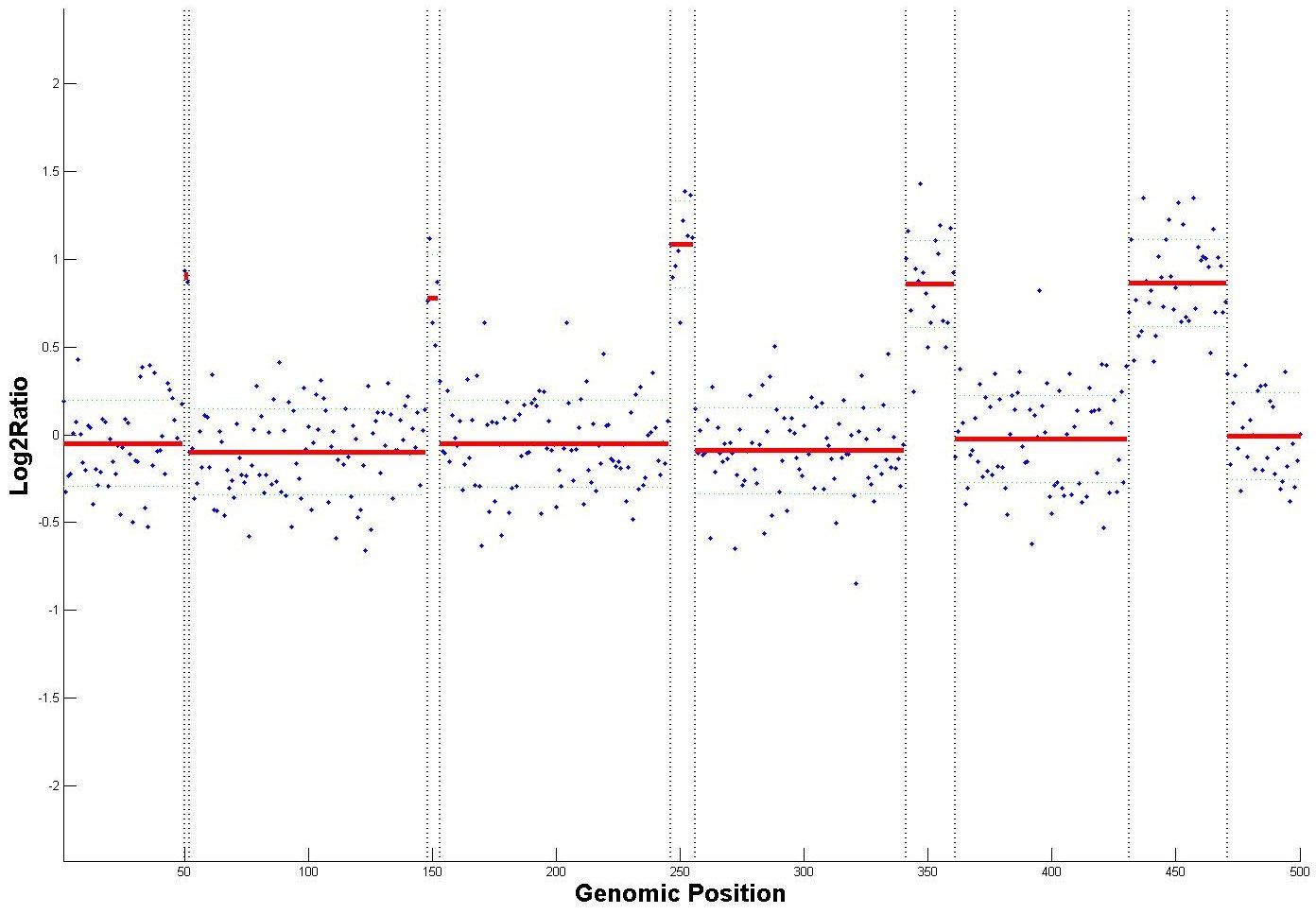}}
  \end{center}
  \caption{Performance of \trimmer, \dnacopy, and \cghseg\ on  denoising synthetic aCGH data from \cite{1181383}. \trimmer\ and \cghseg\ 
exhibit excellent precision and recall whereas \dnacopy\ misses two consecutive genomic positions with DNA copy number equal to 3.}
  \label{fig:laidataset}
\end{figure*}

\subsection{Synthetic Data} 

We use the synthetic data published in \cite{1181383}. 
The data consist of five aberrations of increasing widths
of 2, 5, 10, 20 and 40 probes, respectively, with Gaussian noise N(0,0.25$^2$).  
Figure~\ref{fig:laidataset} shows the performance of \trimmer, \dnacopy, and \cghseg. 
Both \trimmer\ and \cghseg\ correctly detect all aberrations, while \dnacopy\ misses the first, smallest region.
Run time for \trimmer\ is 0.007 sec, compared to 1.23 sec for \cghseg\ and 60 sec for \dnacopy.

\subsection{Coriell Cell Lines}

\begin{table*}[!htb]
\centering
\begin{tabular} {|c|c|c|c|} \hline
Cell Line/Chromosome & \trimmer\ & \dnacopy\ & \cghseg\ \\ \hline
          
GM03563/3     &     \checkmark         & \checkmark    &  \checkmark \\
GM03563/9     &     \checkmark         &    No         &  \checkmark  \\
GM03563/False &        -               &    -          &  - \\ \hline

GM00143/18     &     \checkmark        &   \checkmark   &\checkmark \\
GM00143/False  &      -                &       -      & - \\ \hline
          
GM05296/10 &    \checkmark  & \checkmark  &  \checkmark \\
GM05296/11 &     \checkmark    &  \checkmark  &  \checkmark  \\
GM05296/False &  -   & - & 4,8 \\ \hline
          
GM07408/20 &   \checkmark   &  \checkmark  &  \checkmark \\
GM07408/False &    -   &  - & - \\ \hline 

GM01750/9 &   \checkmark    & \checkmark  &  \checkmark \\
GM01750/14 &   \checkmark    & \checkmark  &  \checkmark  \\
GM01750/False &   -   &  - & - \\ \hline 

GM03134/8 &    \checkmark    & \checkmark  &  \checkmark  \\
GM03134/False &  -  &  - & 1 \\ \hline 
  
GM13330/1 &   \checkmark    &  \checkmark  &   \checkmark  \\
GM13330/4 &   \checkmark    &  \checkmark  &   \checkmark   \\
GM13330/False &   -         &     -        &  -    \\\hline
          
GM03576/2 &   \checkmark    & \checkmark   &  \checkmark \\
GM03576/21 &   \checkmark   & \checkmark   &  \checkmark \\
GM03576/False &   -         &      -       &  - \\\hline
          
GM01535/5 &    \checkmark   &   \checkmark  & \checkmark \\
GM01535/12 &    \checkmark  &    No         & \checkmark \\
GM01535/False & -           &   -           & 8 \\\hline
          
GM07081/7 &    \checkmark   &   \checkmark & \checkmark \\
GM07081/15 &    No   &   No & No \\
GM07081/False &  -   & - & 11 \\\hline
          
GM02948/13 &      \checkmark   &  \checkmark  &  \checkmark  \\
GM02948/False &  7    & 1 &  2 \\ \hline

GM04435/16 &    \checkmark   &  \checkmark  &  \checkmark  \\
GM04435/21 &     \checkmark  &  \checkmark  & \checkmark \\
GM04435/False &  -  & - & 8,17 \\\hline
          
GM10315/22 &     \checkmark   &  \checkmark  & \checkmark\\
GM10315/False &  -    & - & - \\ \hline 

GM13031/17 &       \checkmark  &  \checkmark  & \checkmark  \\
GM13031/False &    -   &  -  &   -  \\ \hline
         
GM01524/6 &        \checkmark   &  \checkmark  & \checkmark  \\
GM01524/False &     -  &  -  &- \\  \hline
\end{tabular}
\caption{Results from applying \trimmer, \dnacopy, and \cghseg\ to 15 cell lines. Rows with listed chromosome numbers (e.g., GM03563/3) corresponded to known gains or losses and are annotated with a check mark if the expected gain or loss was detected or a ``No'' if it was not.  Additional rows list chromosomes on which segments not annotated in the benchmark were detected; we presume these to be false positives.}
\label{tab:coriellevaluation}
\end{table*}

The first real dataset we use to evaluate our method is the Coriell cell line BAC
array CGH data \cite{coriell}, which is widely considered a ``gold standard'' 
dataset. The dataset is derived from 15 fibroblast cell lines 
using the normalized average of 
$\log_2$ fluorescence relative to a diploid reference.
To call gains or losses of inferred segments, we assign each segment the mean intensity of its probes and then apply a simple threshold test to determine if the mean is abnormal.  We follow \cite{thresholds} in favoring $\pm$0.3 out of the wide variety of thresholds that have been used \cite{thresholds2}.

Table~\ref{tab:coriellevaluation} summarizes  the performance of \trimmer, \dnacopy\ and \cghseg\ relative to previously annotated gains and losses in the Corielle dataset. The table shows notably better performance for \trimmer\ compared to either alternative method.  \trimmer\ finds 22 of 23 expected segments with one false positive.  \dnacopy\ finds 20 of 23 expected segments with one false positive.  \cghseg\ finds 22 of 23 expected segments with seven false positives.  \trimmer\ thus achieves the same recall as \cghseg\ while outperforming it in precision and the same precision as \dnacopy\ while outperforming it in recall.
In cell line GM03563, \dnacopy\ fails to detect a region of two points
which have undergone a loss along chromosome 9, in accordance with the results 
obtained using the Lai et al. \cite{1181383} synthetic data. 
In cell line GM03134, \cghseg\ makes a false positive along chromosome 1
which both \trimmer\ and \dnacopy\ avoid. 
In cell line GM01535, \cghseg\ makes a false positive along chromosome 8
and \dnacopy\ misses the aberration along chromosome 12.
\trimmer, however, performs ideally on this cell line. 
In cell line GM02948, \trimmer\ makes a false positive along chromosome 7, finding
a one-point segment in 7q21.3d at genomic position 97000 
whose value is equal to 0.732726. All other methods also make false positive errors
on this cell line.  In GM7081, all three methods fail to find an annotated aberration on chromosome 15.  In addition, \cghseg\ finds a false positive on chrosome 11.

\trimmer\ also substantially outperforms the comparative methods in
run time, requiring 5.78 sec for the full data set versus 8.15 min for
\cghseg\ (an 84.6-fold speedup) and 47.7 min for \dnacopy\ (a 495-fold
speedup).

\subsection{Breast Cancer Cell Lines}

To further illustrate the performance of \trimmer\ and compare it to \dnacopy\ and \cghseg, we applied it to the Berkeley Breast Cancer cell line database \cite{berkeley}.
The dataset consists of 53 breast cancer cell lines that capture most of the recurrent genomic and 
transcriptional characteristics of 145 primary breast cancer cases. We do not have an accepted ``answer key'' for this data set, but it provides a more extensive basis for detailed comparison of differences in performance of the methods on common data sets, as well as an opportunity for novel discovery.  While we have applied the methods to all chromosomes in all cell lines, space limitations prevent us presenting the full results here.  We therefore arbitrarily selected three of the 53 cell lines and selected three chromosomes per cell line that we believed would best illustrate the comparative performance of the methods.  The Genes-to-Systems Breast Cancer
Database \url{http://www.itb.cnr.it/breastcancer}~\cite{G2SBCD} was used to identify known breast cancer markers in regions predicted to be gained or lost by one or more of the methods, with the UCSC Genome Browser
\url{http://genome.ucsc.edu/} used to verify the placement of genes.
 
We note that \trimmer\ again had a substantial advantage in run time.  For the full data set, \trimmer\ required 22.76 sec, compared to 23.3 min for \cghseg\ (a 61.5-fold increase), and 4.95 hrs for \dnacopy\ (a 783-fold increase).

\begin{figure*}
		\begin{tabular}{ccc} 
		\includegraphics[width=0.33\textwidth]{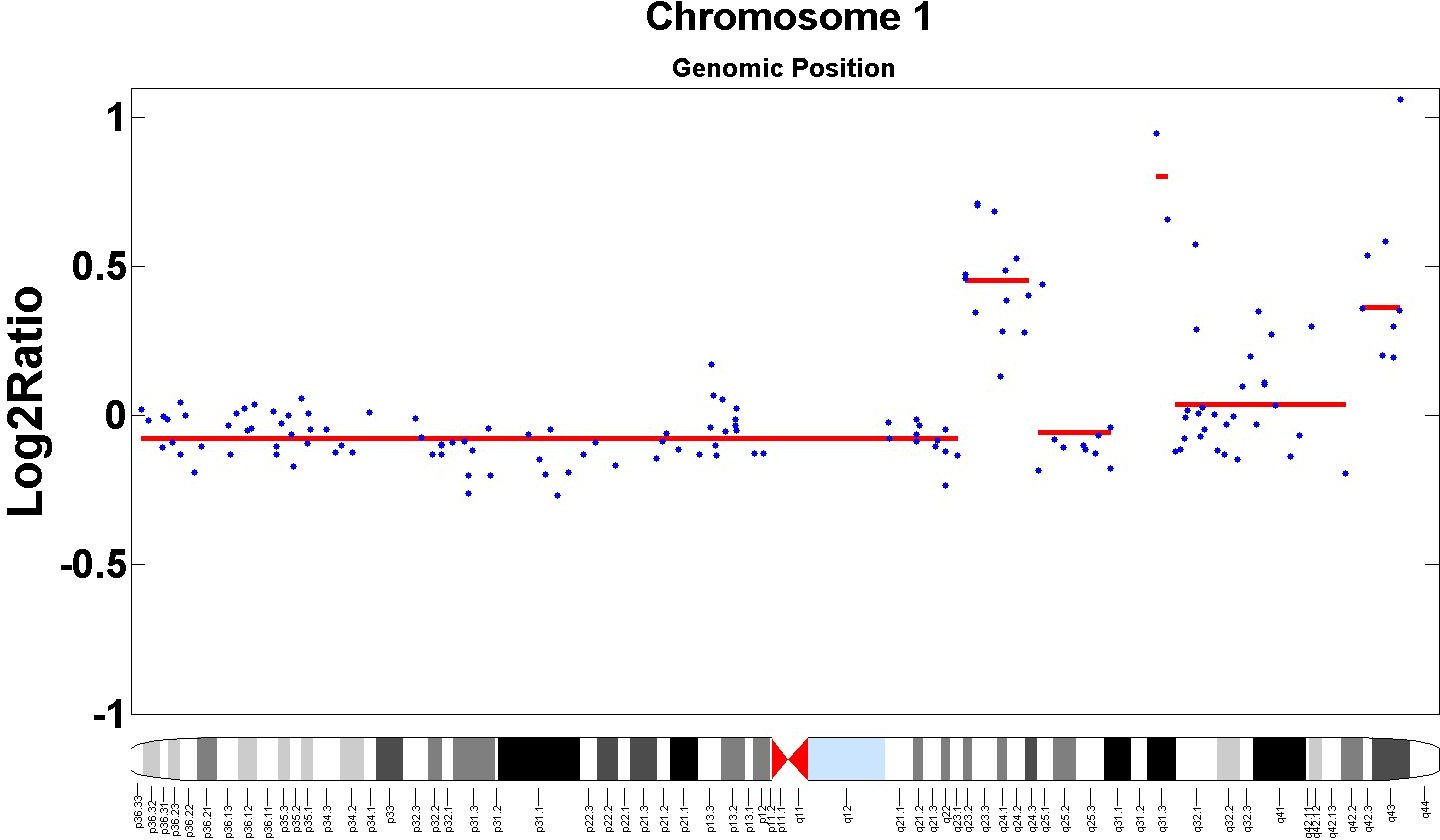} & \includegraphics[width=0.33\textwidth]{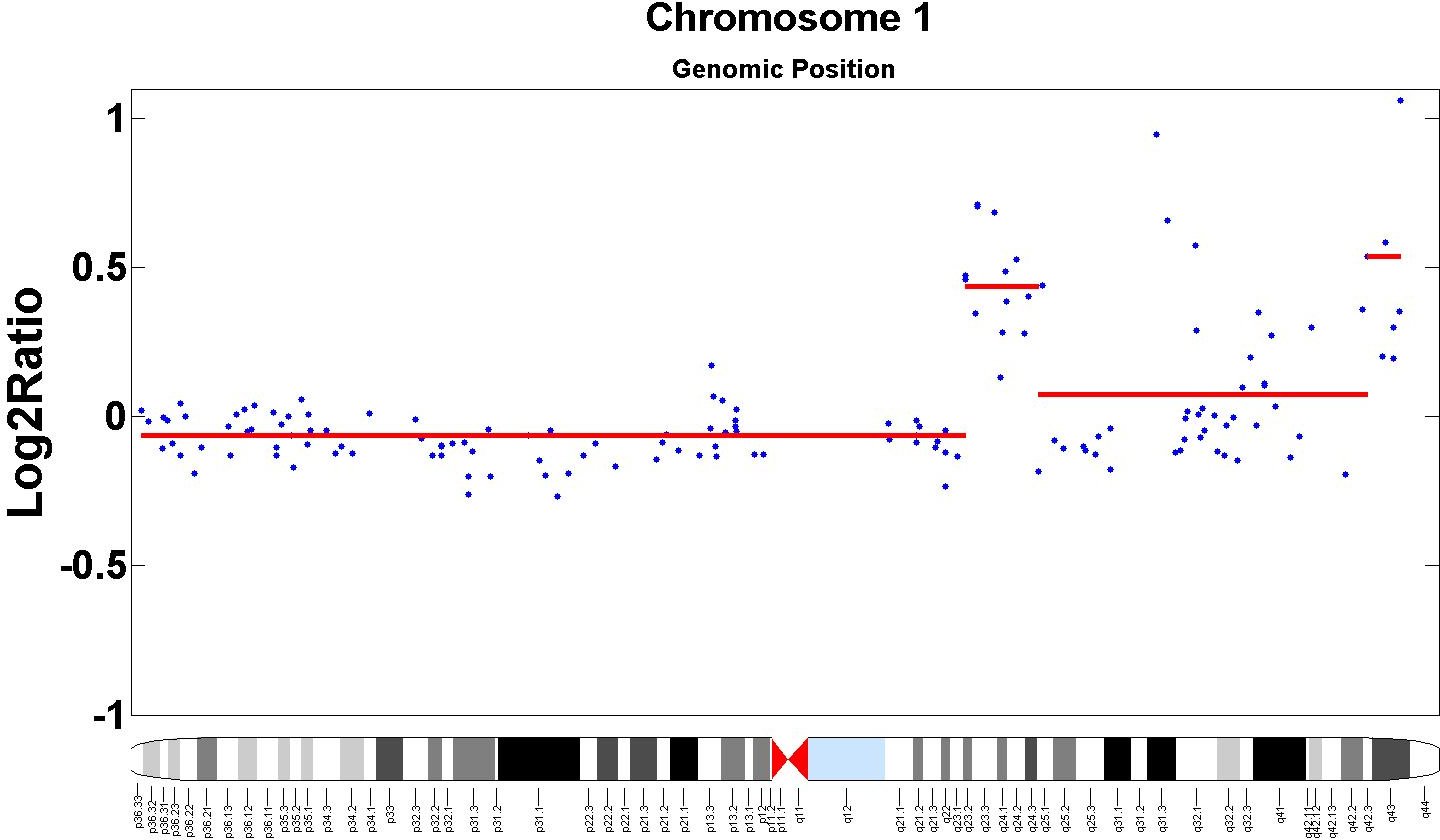} &    \includegraphics[width=0.33\textwidth]{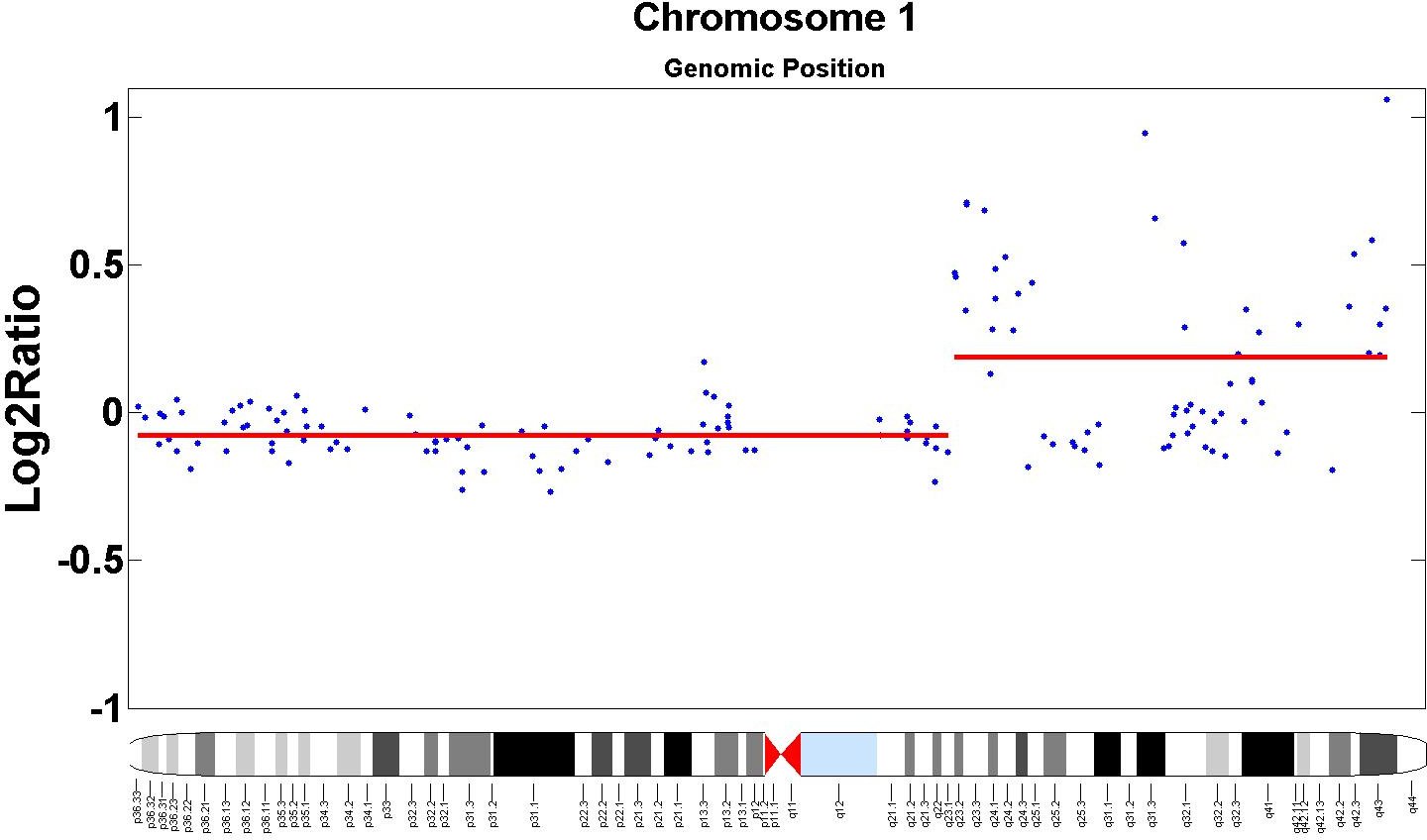}   \\
		(a) &(b) &(c) \\
		\includegraphics[width=0.33\textwidth]{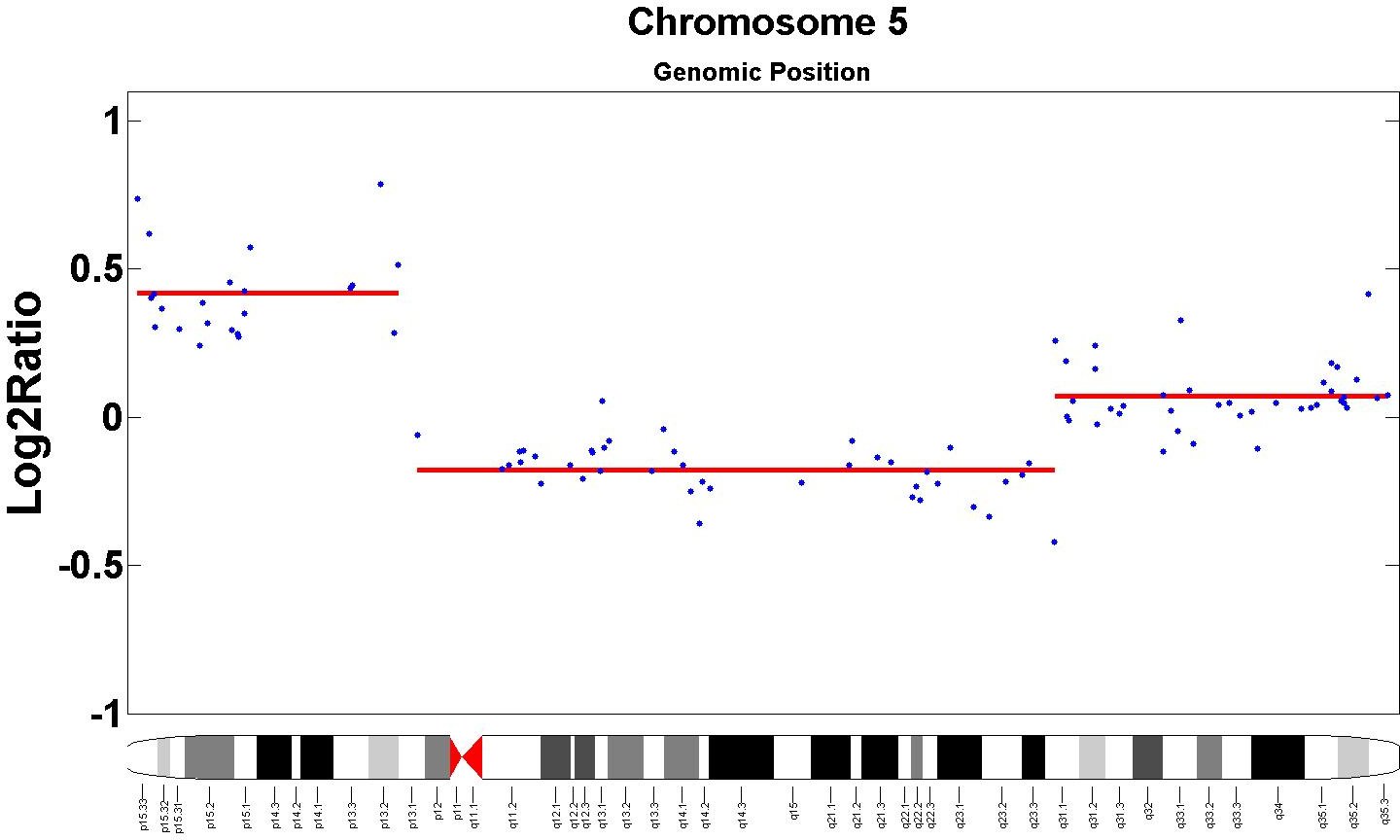} &		\includegraphics[width=0.33\textwidth]{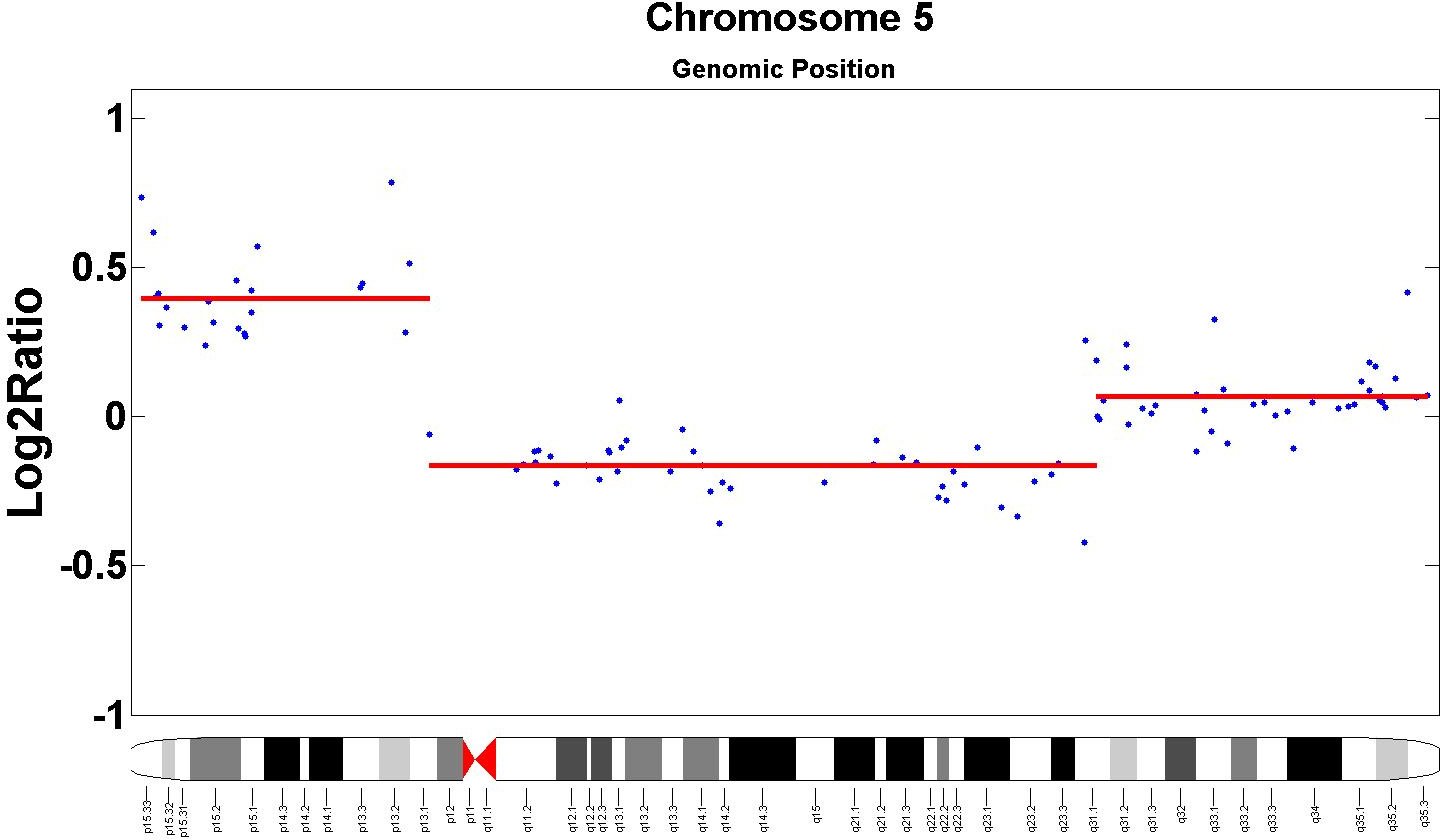} &    \includegraphics[width=0.33\textwidth]{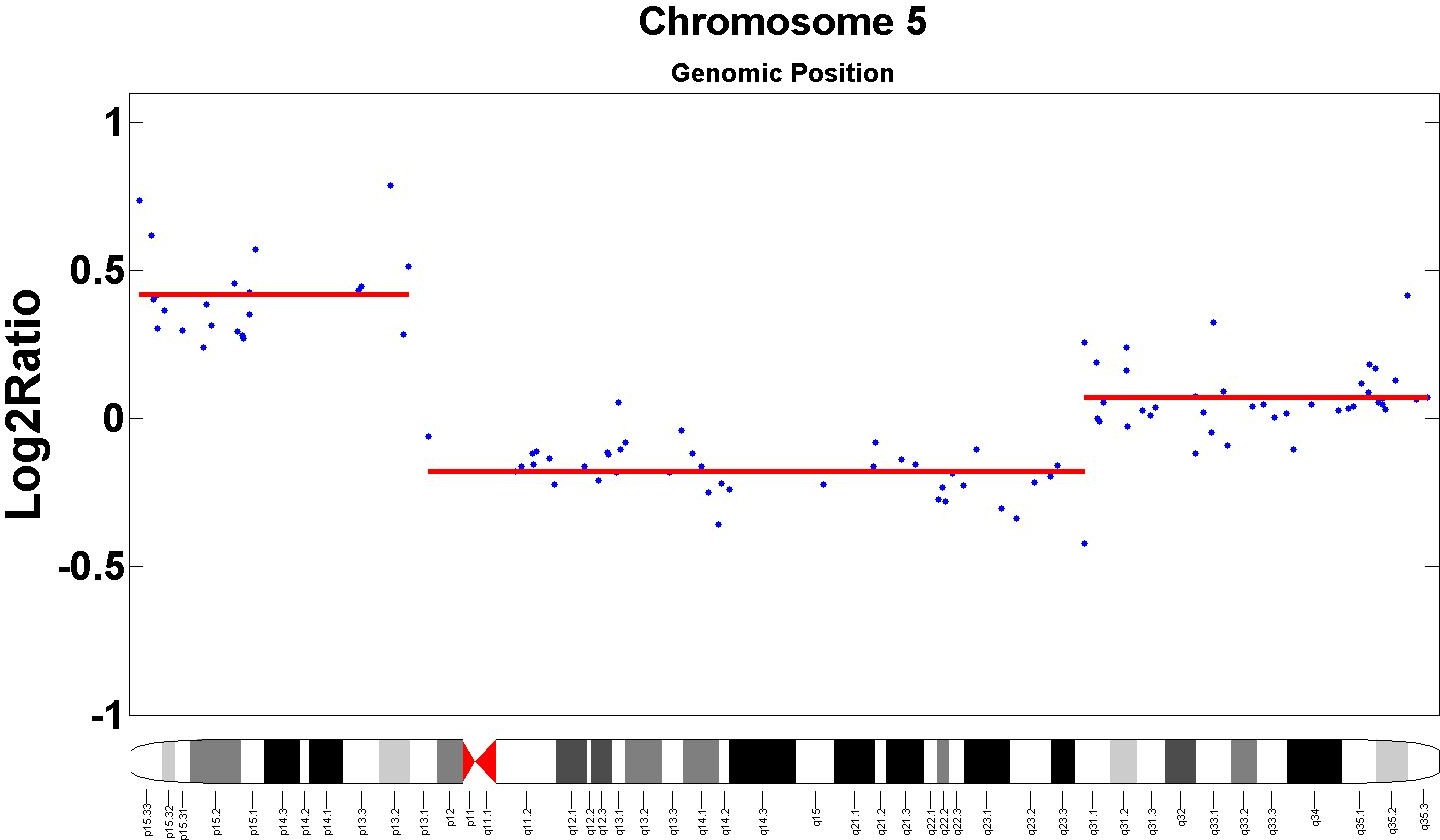}   \\
		(d) &(e) &(f) \\
		\includegraphics[width=0.33\textwidth]{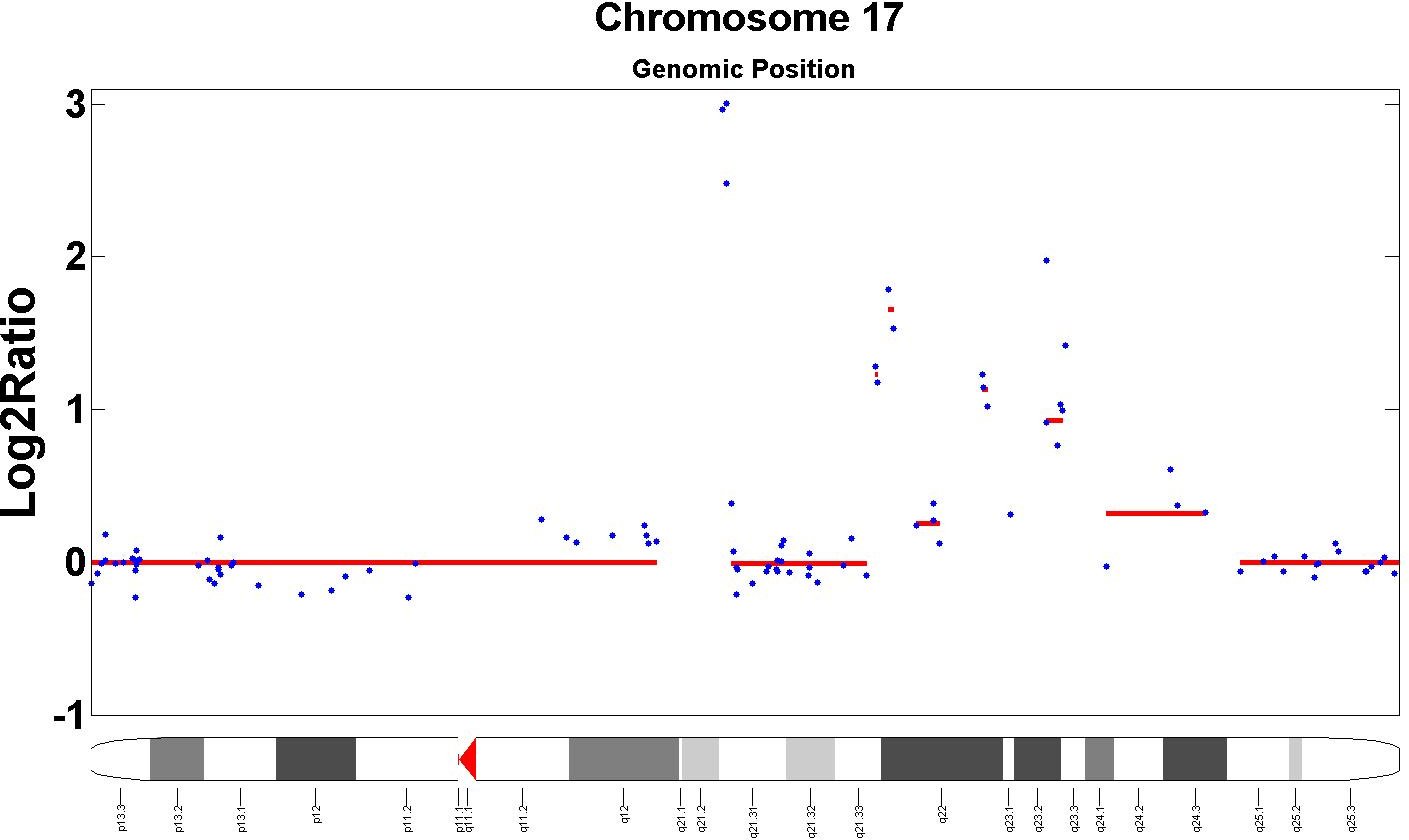} & \includegraphics[width=.33\textwidth]{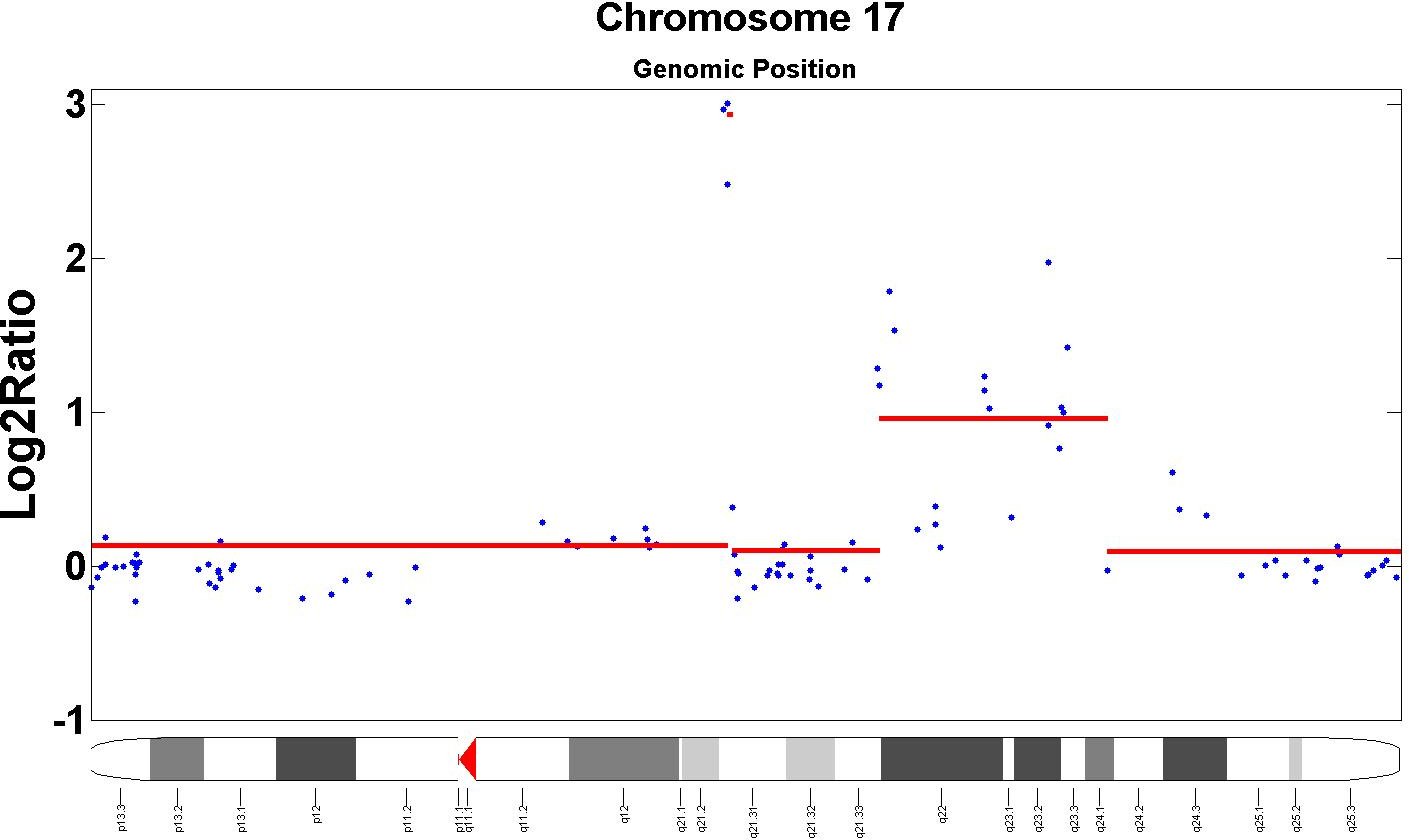} &    \includegraphics[width=0.33\textwidth]{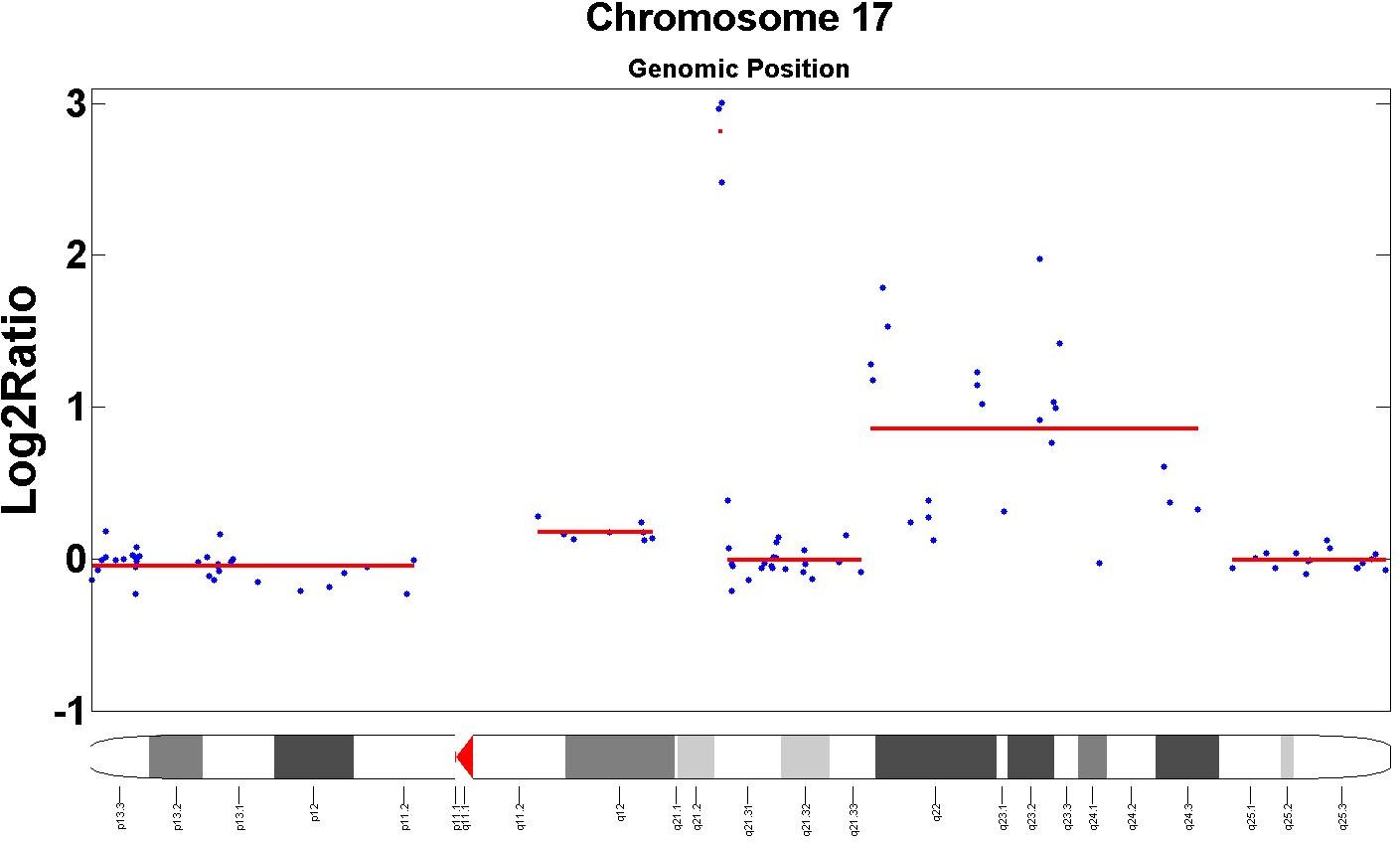}  \\
		(g) &(h) &(i) \\
		\end{tabular}
\caption{Visualization of segmentation output of \trimmer, \dnacopy, and \cghseg\ for cell line BT474 on chromosomes 1 (a,b,c), 5 (d,e,f), and 17 (g,h,i).  (a,d,g) \trimmer output.  (b,e,h) \dnacopy output. (c,f,i) \cghseg output.}
\label{fig:bt474}
\end{figure*}

\paragraph{\textbf{Cell Line BT474:}}
Figure~\ref{fig:bt474} shows the performance of each method on the BT474 cell line.  The three methods report different results for chromsome 1, 
as shown in Figures~\ref{fig:bt474}(a,b,c), with all three detecting amplification in the q-arm but differing in the detail of resolution.  
\trimmer\ is the only method that detects region 1q31.2-1q31.3 as aberrant. 
This regions hosts gene NEK7, a candidate oncogene \cite{nek7} and gene KIF14, 
a predictor of grade and outcome in breast cancer \cite{KIF14}.
\trimmer\ and \dnacopy\ annotate the region 1q23.3-1q24.3 as
amplified. This region hosts several genes previously implicated in breast
cancers~\cite{G2SBCD}, such 
as CREG1 (1q24), POU2F1 (1q22-23), RCSD1 (1q22-q24), and BLZF1 (1q24).
Finally, \trimmer\ alone reports independent amplification of the gene CHRM3, a a marker of metastasis in breast cancer patients~\cite{G2SBCD}.

For chromosome 5 (Figures~\ref{fig:bt474}(d,e,f)), the behavior of the
three methods is almost identical, with all three reporting
amplification of a region known to contain many breast cancer markers,
including MRPL36 (5p33), ADAMTS16 (5p15.32), POLS (5p15.31), ADCY2
(5p15.31), CCT5 (5p15.2), TAS2R1 (5p15.31), ROPN1L (5p15.2), DAP
(5p15.2), ANKH (5p15.2), FBXL7 (5p15.1), BASP1 (5p15.1), CDH18
(5p14.3), CDH12 (5p14.3), CDH10 (5p14.2 - 5p14.1), CDH9 (5p14.1) PDZD2
(5p13.3), GOLPH3 (5p13.3), MTMR12 (5p13.3), ADAMTS12 (5p13.3 -
5p13.2), SLC45A2 (5p13.2), TARS (5p13.3), RAD1 (5p13.2), AGXT2
(5p13.2), SKP2 (5p13.2), NIPBL (5p13.2), NUP155 (5p13.2), KRT18P31
(5p13.2), LIFR (5p13.1) and GDNF (5p13.2)~\cite{G2SBCD}.  The only
difference in the assignments is that \dnacopy\ fits one more probe to
this amplified segment.

Finally, for chromosome 17 (Figures~\ref{fig:bt474}(g,h,i)), like chromosome 1, all three detect amplification but \trimmer\ predicts a finer breakdown of the amplified region into independently amplified segments.  All three detect amplification of a region including the major breast cancer biomarkers HER2 (17q21.1) and BRCA1 (17q21) and the additional markers MSI2 (17q23.2) and  TRIM37 (17q23.2)~\cite{G2SBCD}.  While the more discontiguous picture produced by \trimmer\ may appear to be a less parsimonious explanation of the data, a complex combination of fine-scale gains and losses in 17q is in fact well supported by the literature \cite{Orsetti}.

\paragraph{ \textbf{Cell Line HS578T:}}

\begin{figure*}
		\begin{tabular}{ccc} 
		\includegraphics[width=0.33\textwidth]{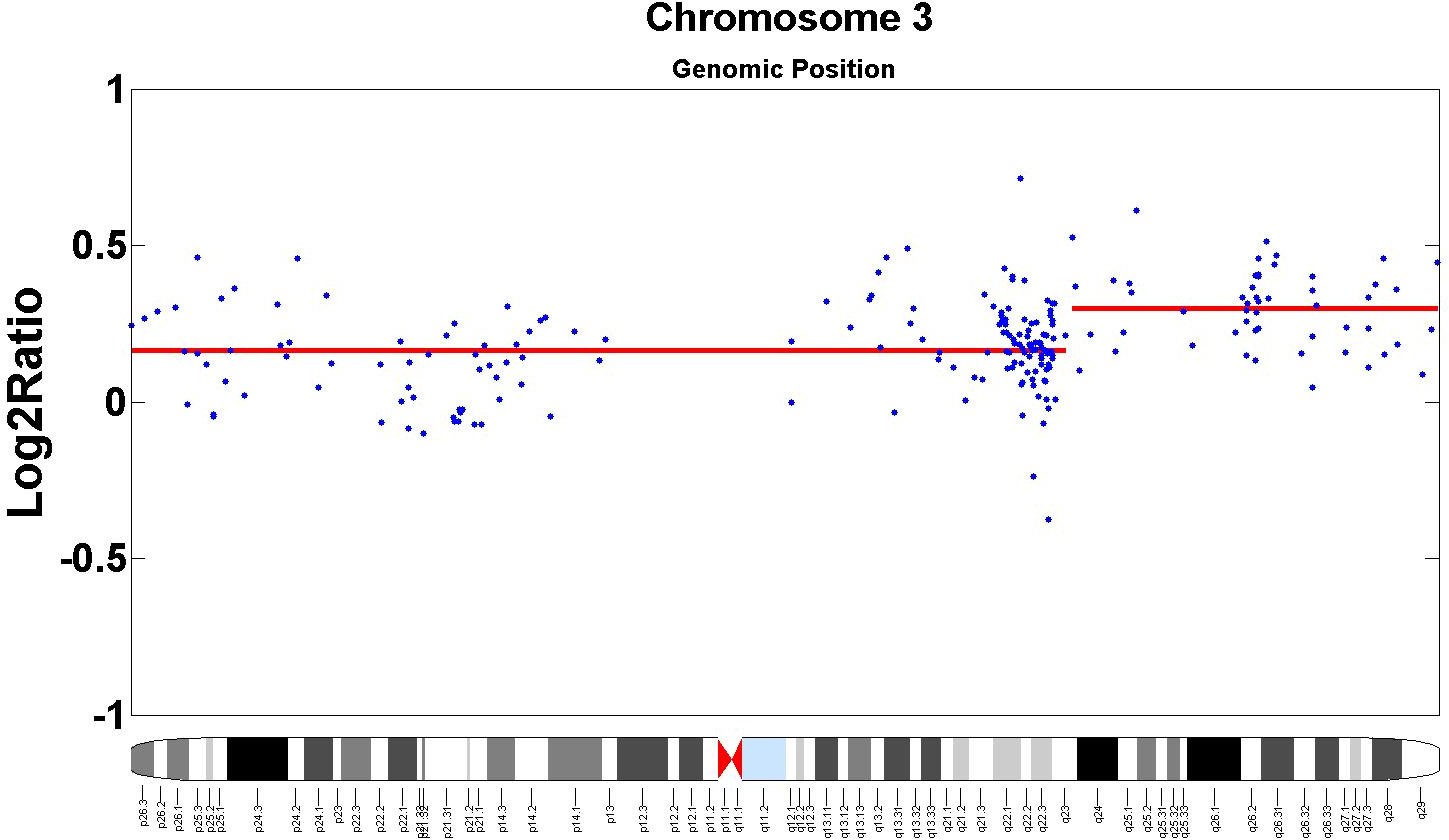} & \includegraphics[width=0.33\textwidth]{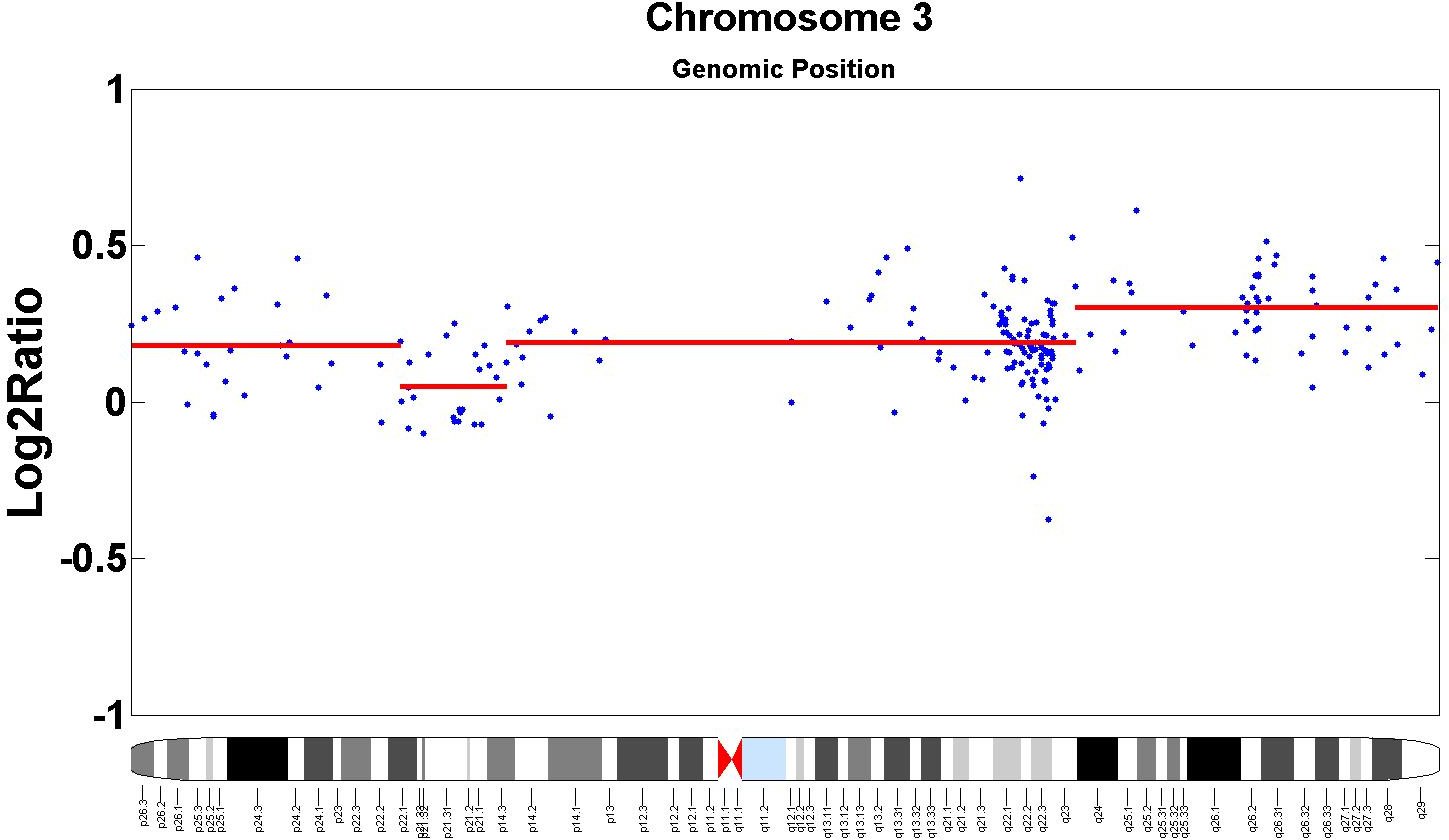} &    \includegraphics[width=0.33\textwidth]{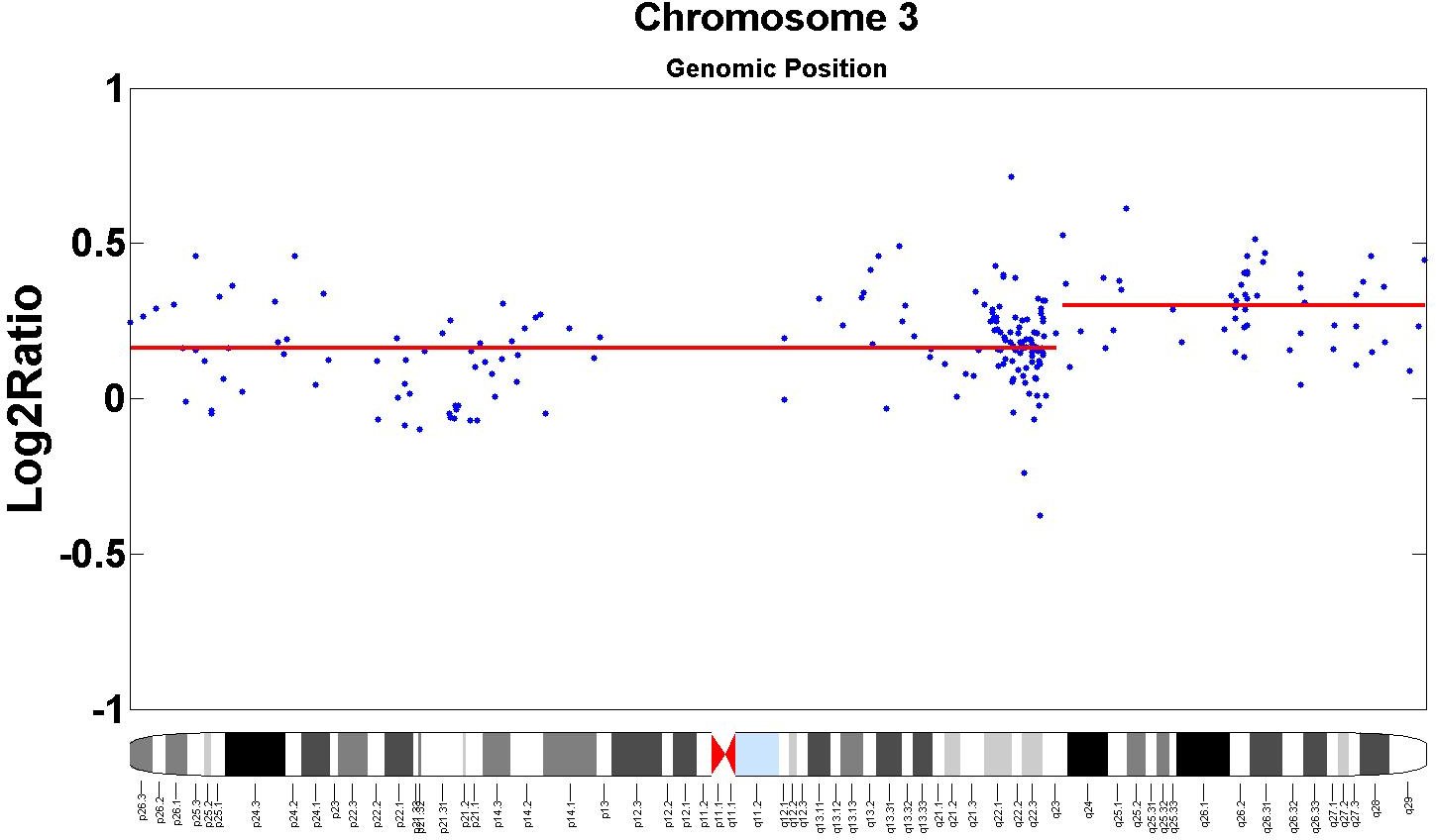} \\
		(a) &(b) &(c) \\
		 \includegraphics[width=0.31\textwidth]{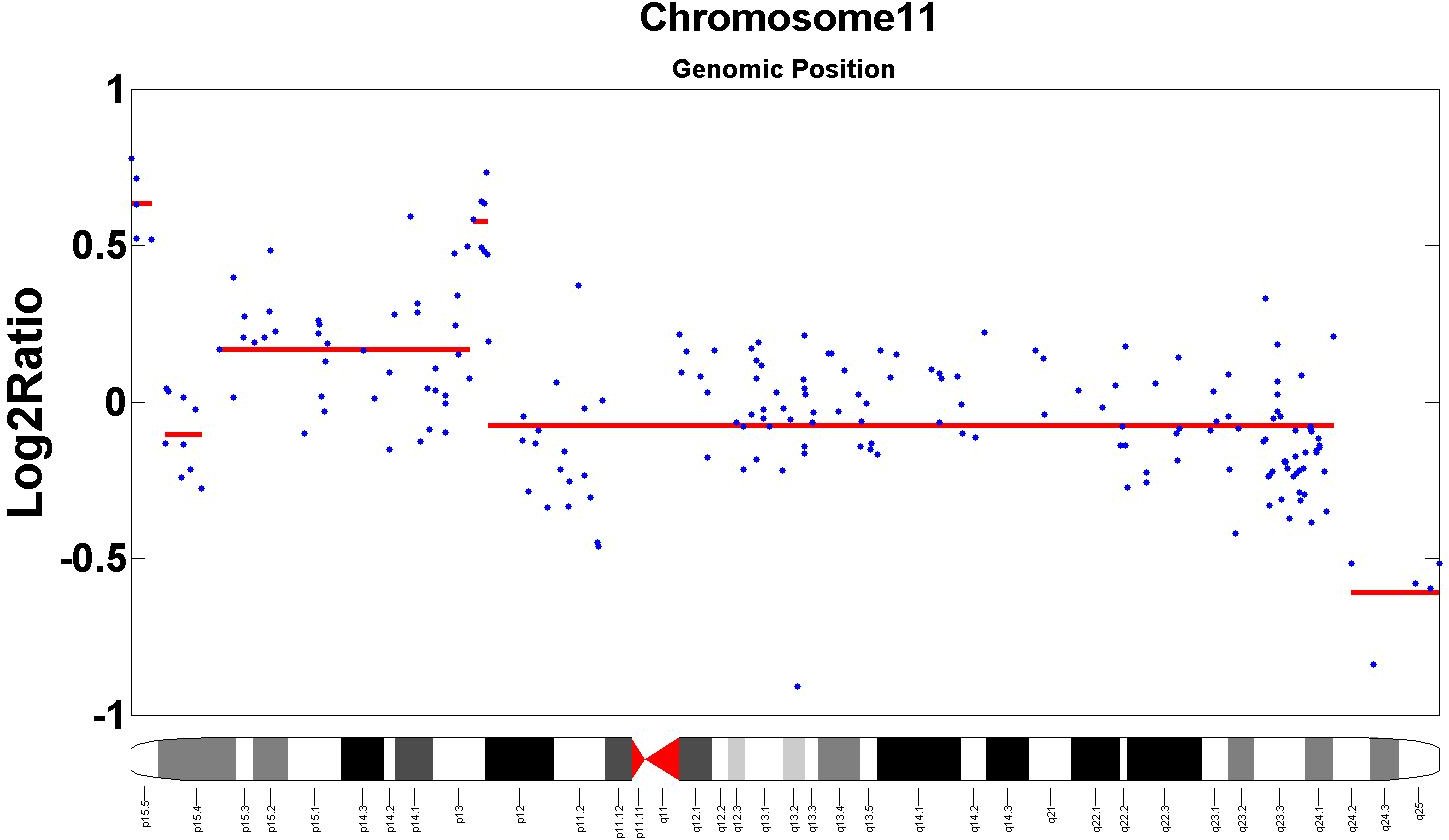} & \includegraphics[width=0.33\textwidth]{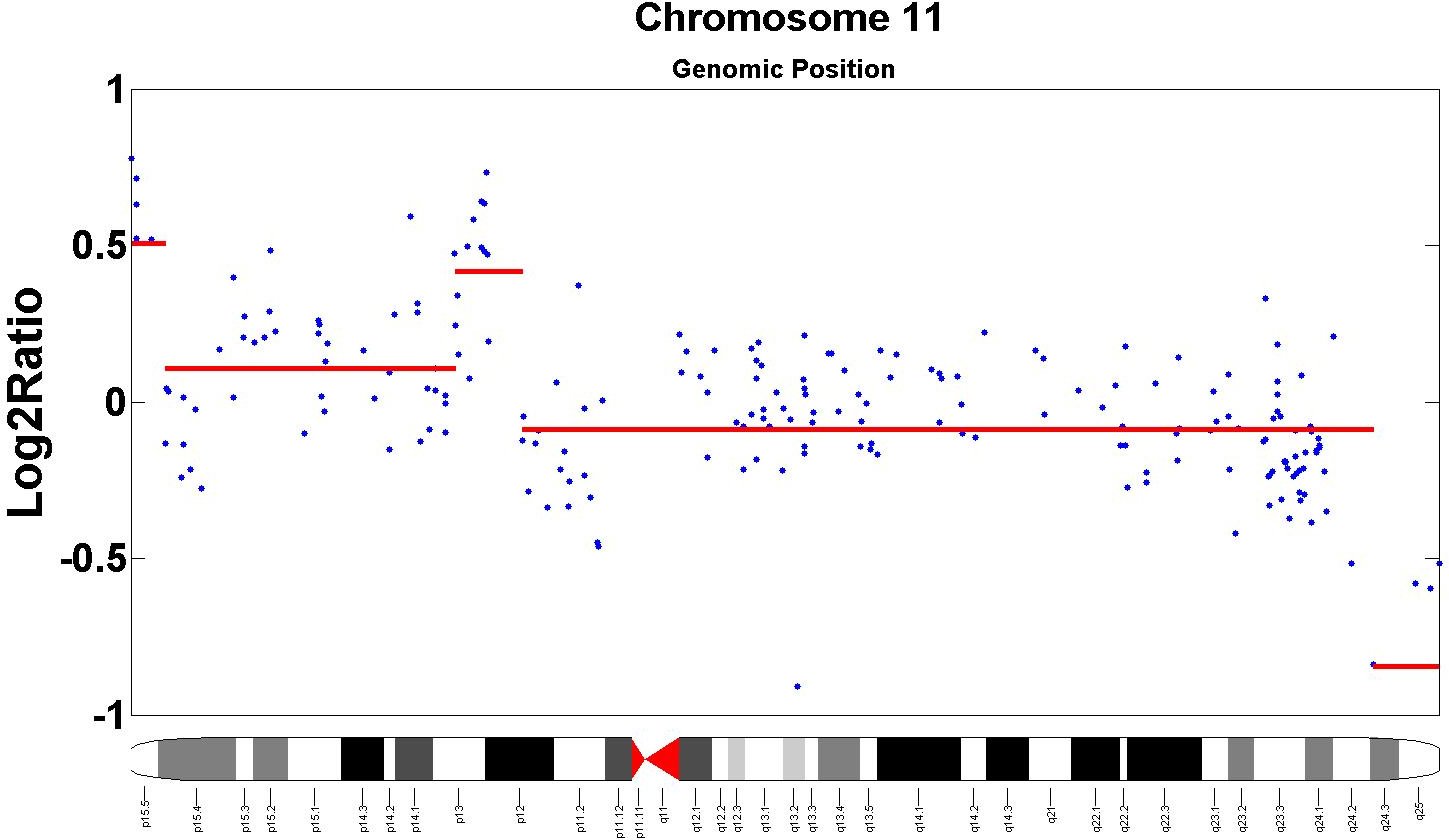} &    \includegraphics[width=0.33\textwidth]{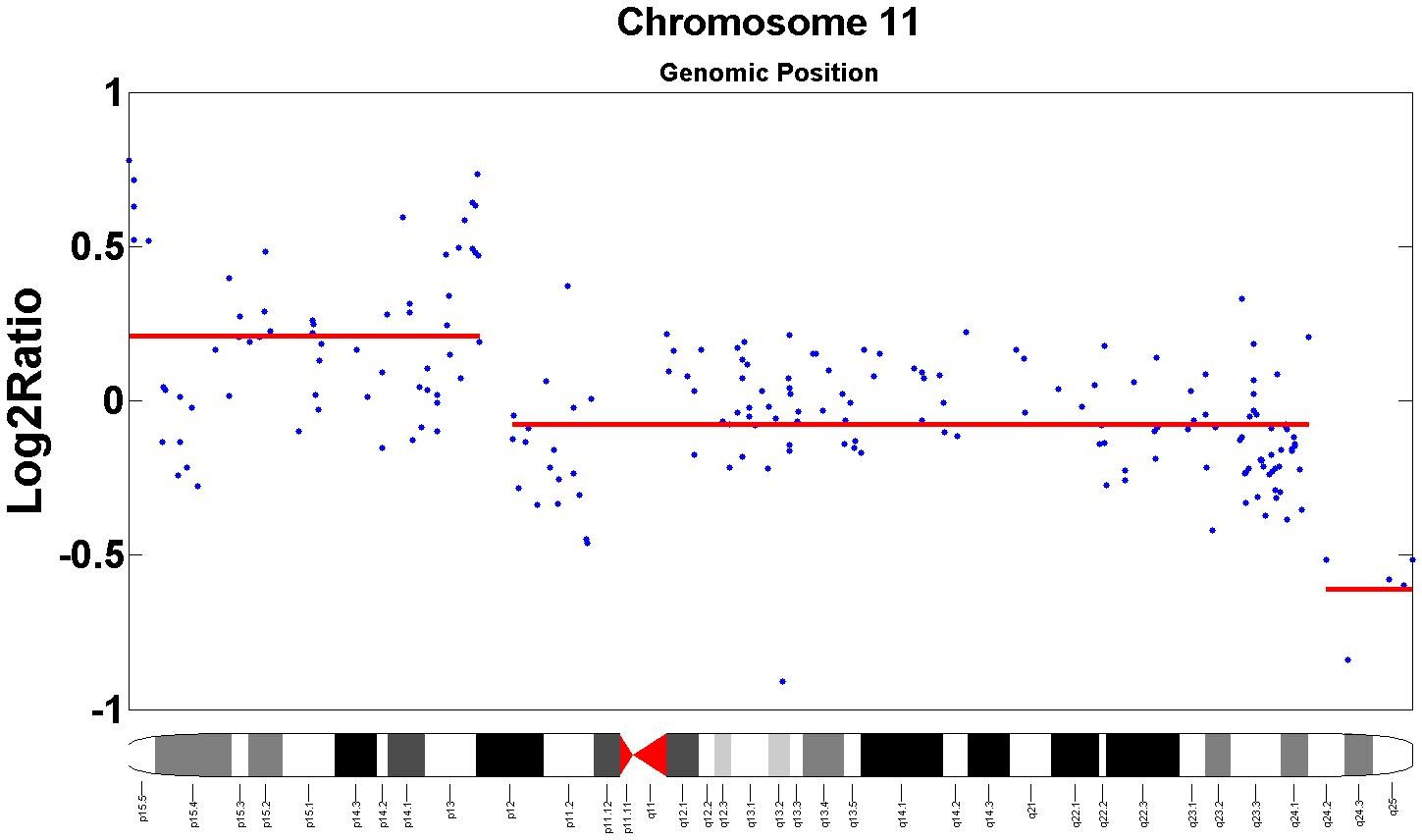} \\
		(d) &(e) &(f) \\
		\includegraphics[width=0.33\textwidth]{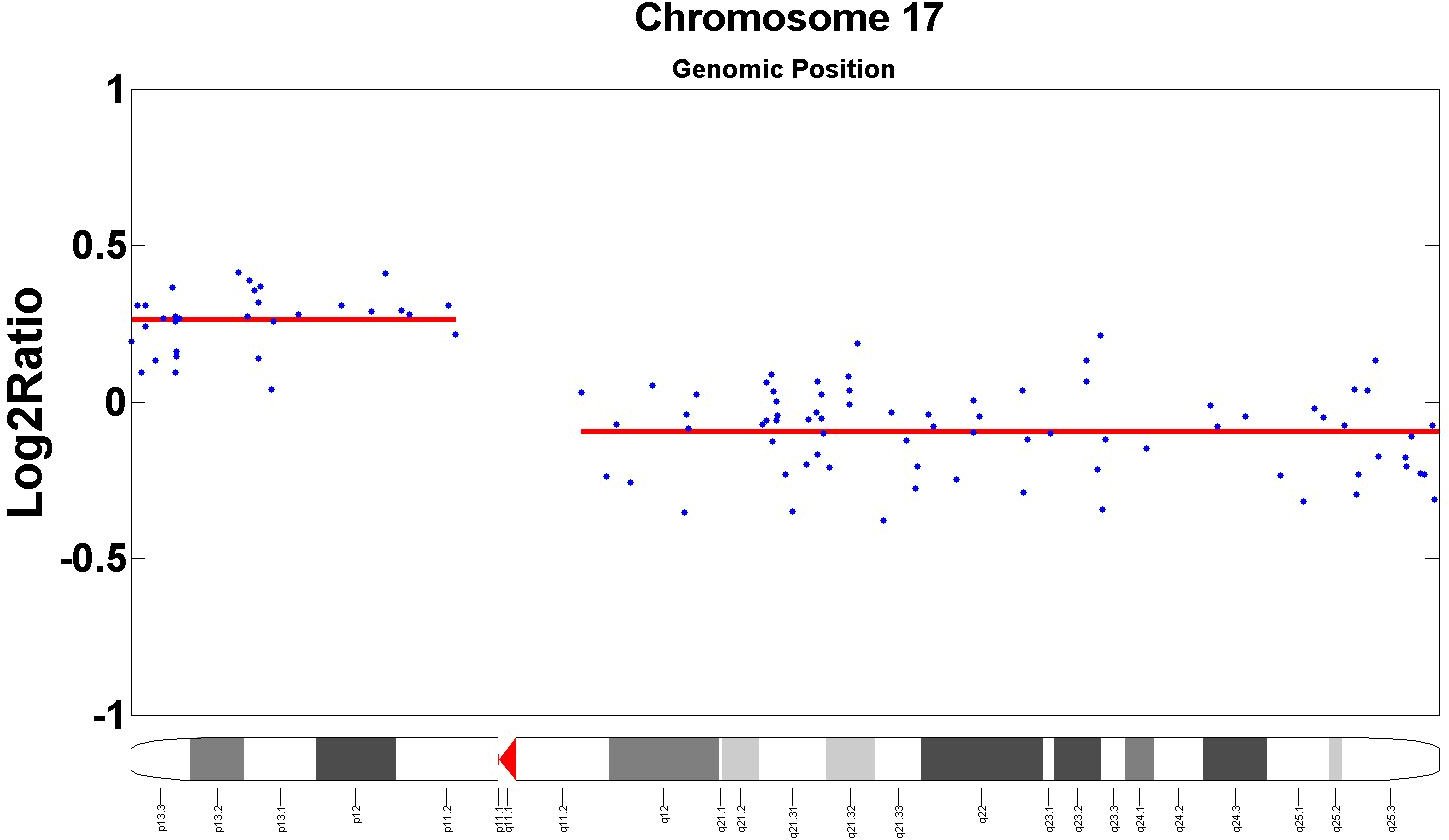} & \includegraphics[width=0.33\textwidth]{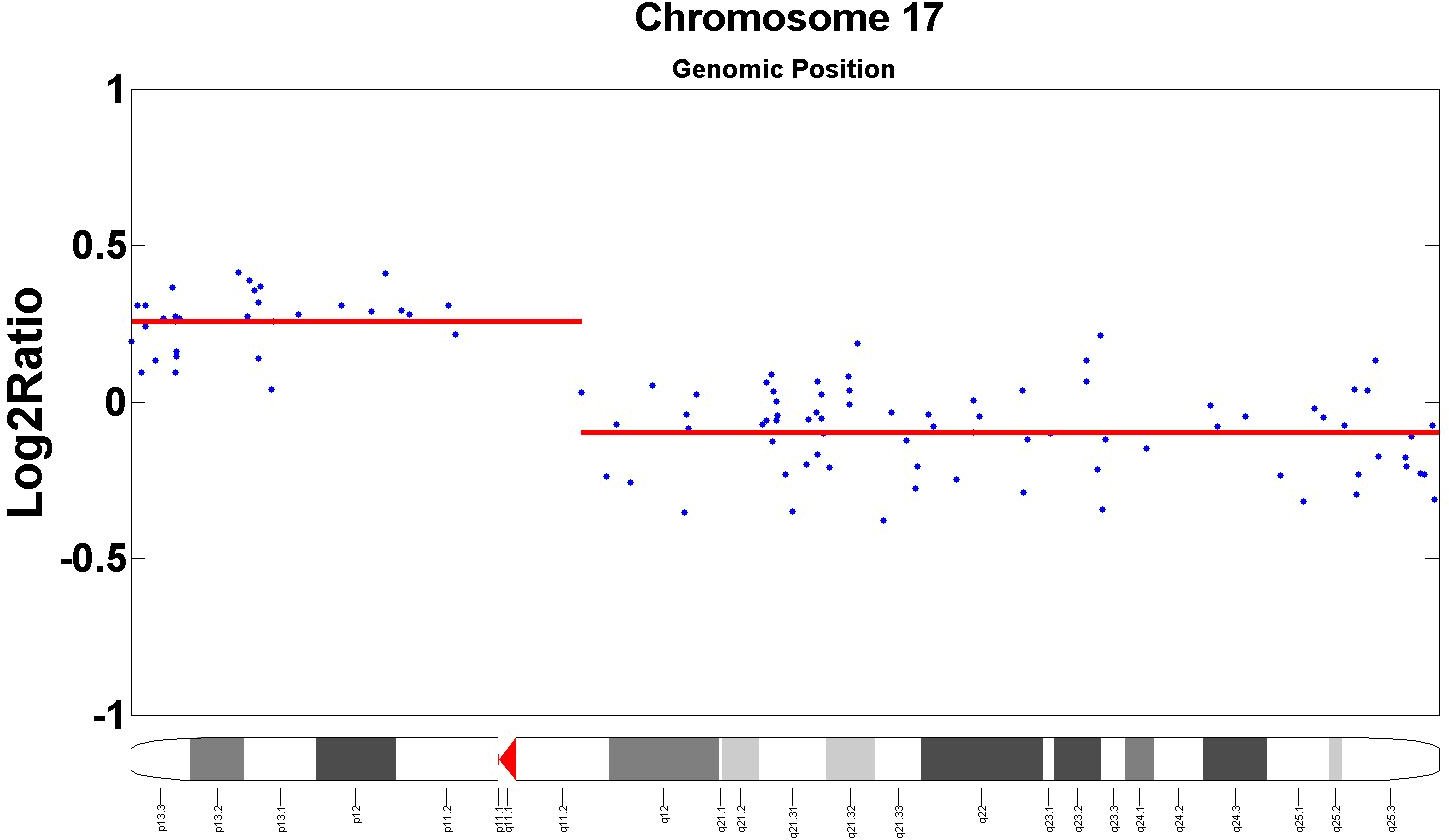} &    \includegraphics[width=0.32\textwidth]{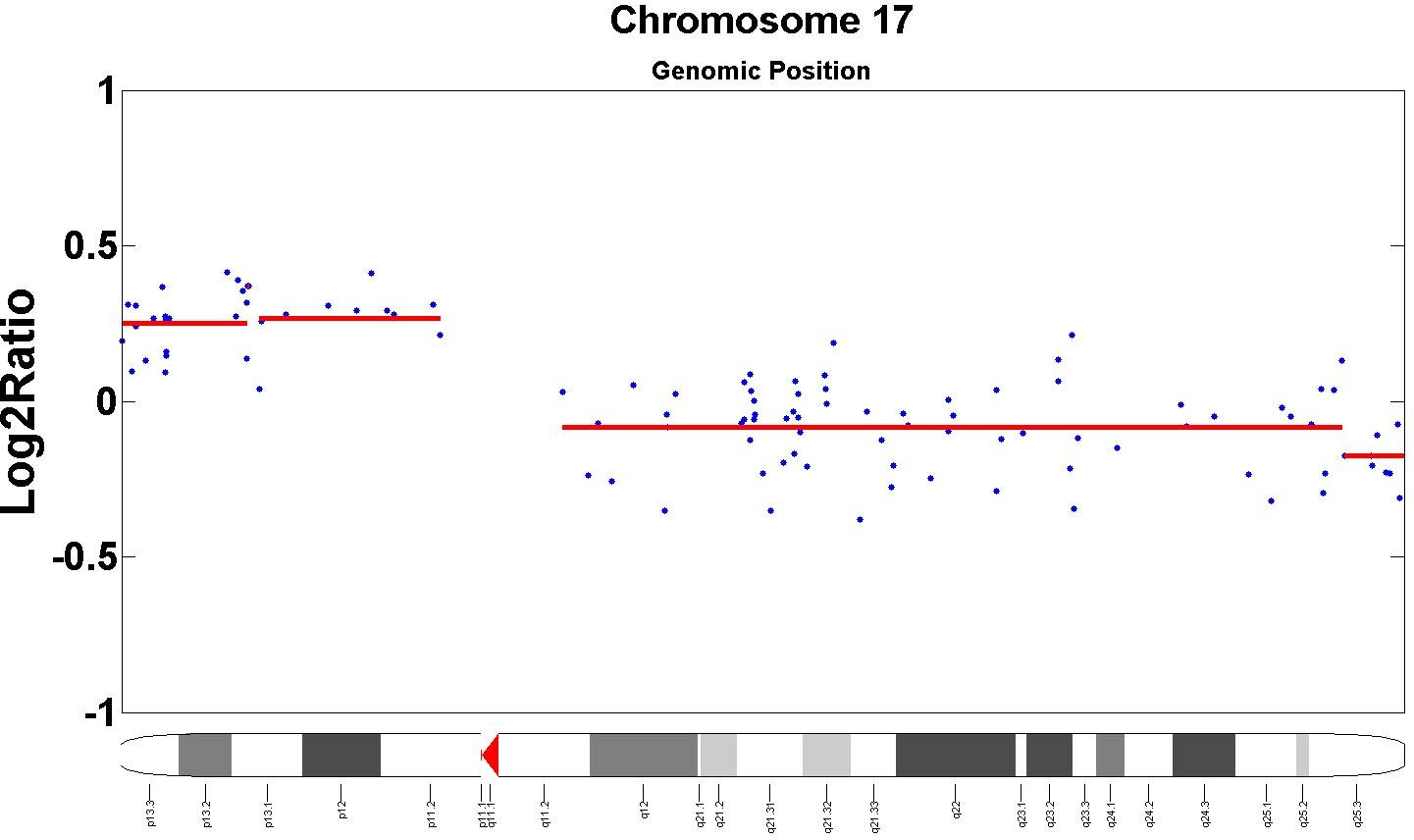} \\
		(g) &(h) &(i) \\
		\end{tabular}
\caption{Visualization of segmentation output of \trimmer, \dnacopy, and \cghseg\ for cell line HS578T on chromosomes 3 (a,b,c), 11 (d,e,f), and 17 (g,h,i).  (a,d,g) \trimmer output.  (b,e,h) \dnacopy output. (c,f,i) \cghseg output.}
\label{fig:HS578T}
\end{figure*}

Figure~\ref{fig:HS578T} compares the methods on cell line HS578T for
chromosomes 3, 11 and 17.  Chromosome 3 (Figures~\ref{fig:HS578T}(a,b,c))
shows identical prediction of an amplification of 3q24-3qter for all
three methods, a region including the key breast cancer markers PIK3CA
(3q26.32)~\cite{pik3ca}, and additional breast-cancer-associated genes
TIG1 (3q25.32), MME (3q25.2), TNFSF10 (3q26), MUC4 (3q29), TFRC
(3q29), DLG1 (3q29)~\cite{G2SBCD}. \trimmer\ and \cghseg\ also have
identical predictions of normal copy number in the p-arm, while
\dnacopy\ reports an additional loss between 3p21 and 3p14.3.  We are
unaware of any known gain or loss in this region associated with
breast cancer.

For chromosome 11 (Figures~\ref{fig:HS578T}(d,e,f)), the methods again
present an identical picture of loss at the q-terminus
(11q24.2-11qter) but detect amplifications of the p-arm at different
levels of resolution.  \trimmer\ and \dnacopy\ detect gain in the
region 11p15.5, the site of the HRAS breast cancer metastasis
marker~\cite{G2SBCD}.  In contrast to \dnacopy, \trimmer\ detects an
adjacent loss region.  While we have no direct evidence this loss is a
true finding, the region of predicted loss does contain EIF3F
(11p15.4), identified as a possible tumor suppressor whose expression
is decreased in most pancreatic cancers and melanomas~\cite{G2SBCD}.
We can thus conjecture that EIF3F is also a tumor suppressor in
breast cancers.

\begin{figure*}
		\begin{tabular}{ccc} 
		\includegraphics[width=0.33\textwidth]{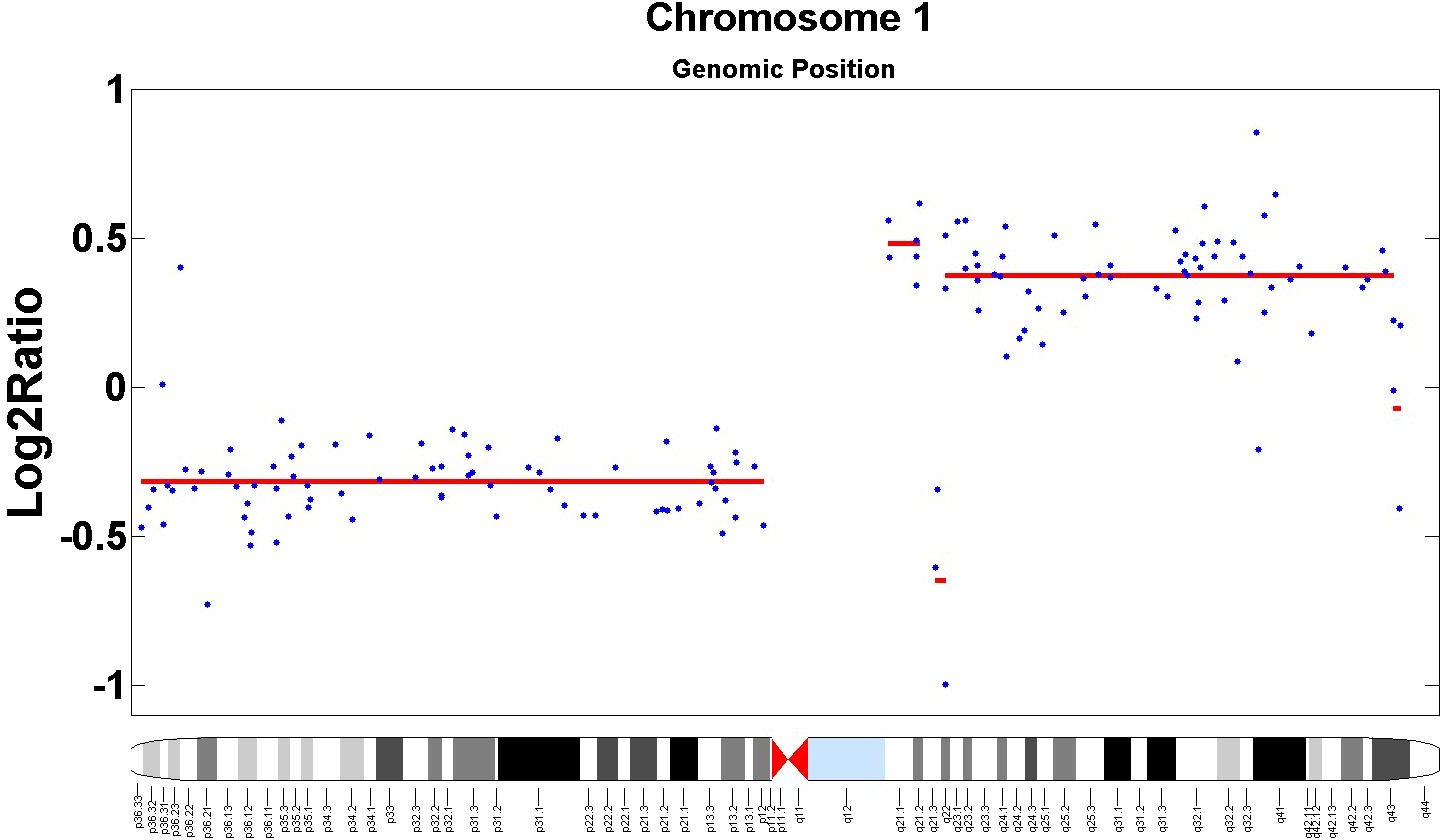} & \includegraphics[width=0.33\textwidth]{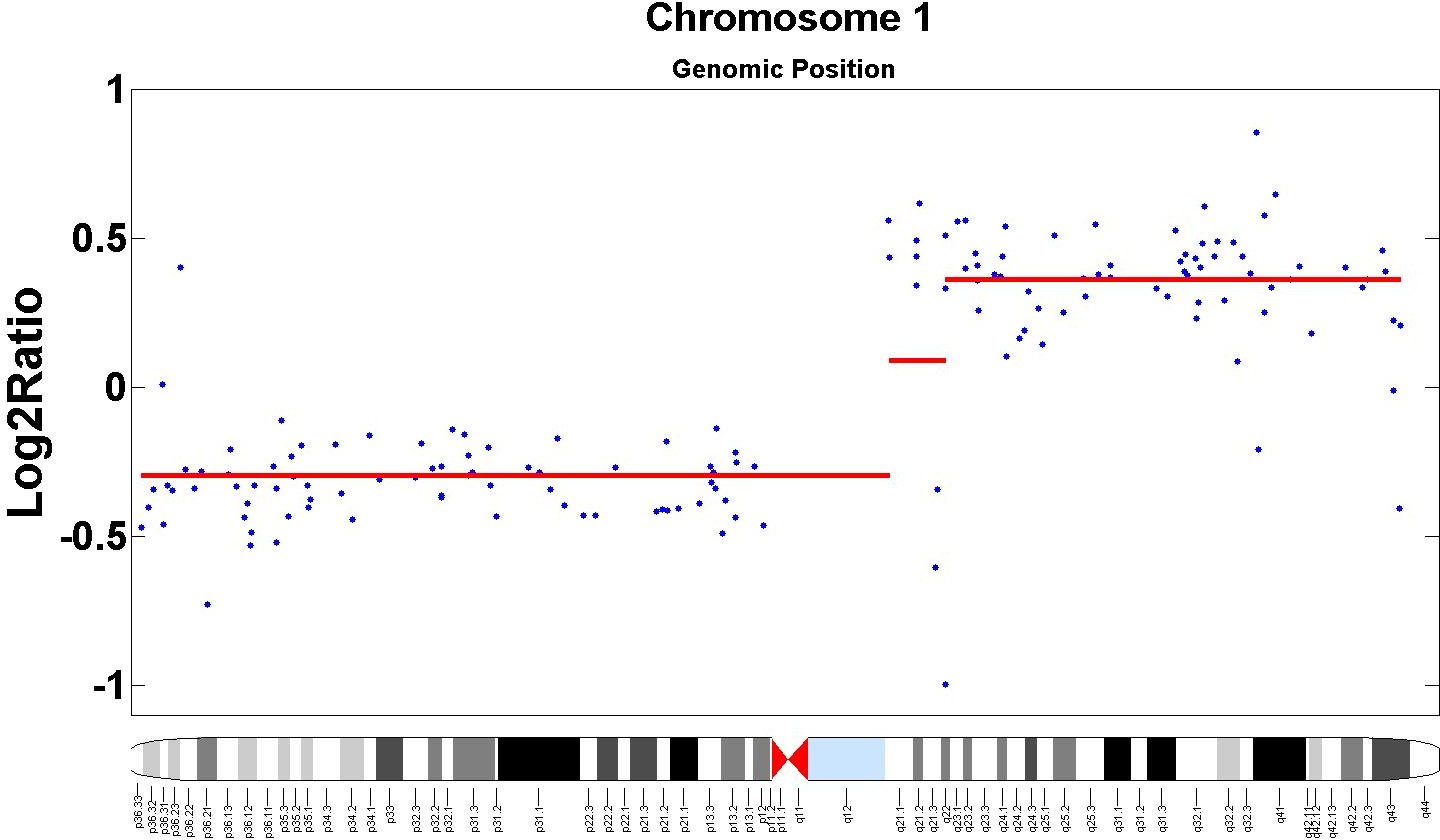} &    \includegraphics[width=0.33\textwidth]{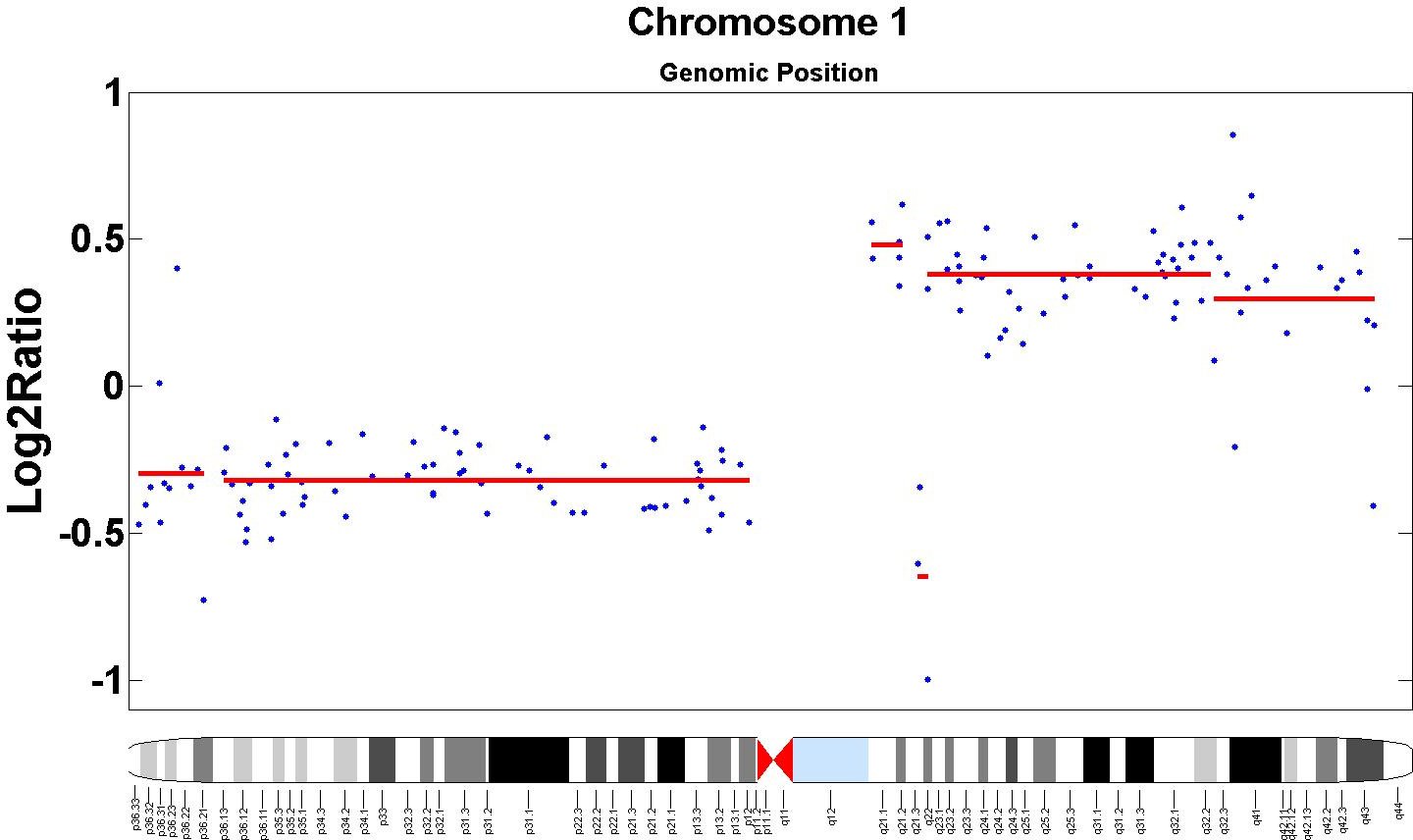}  \\
		(a) &(b) &(c) \\
		\includegraphics[width=0.33\textwidth]{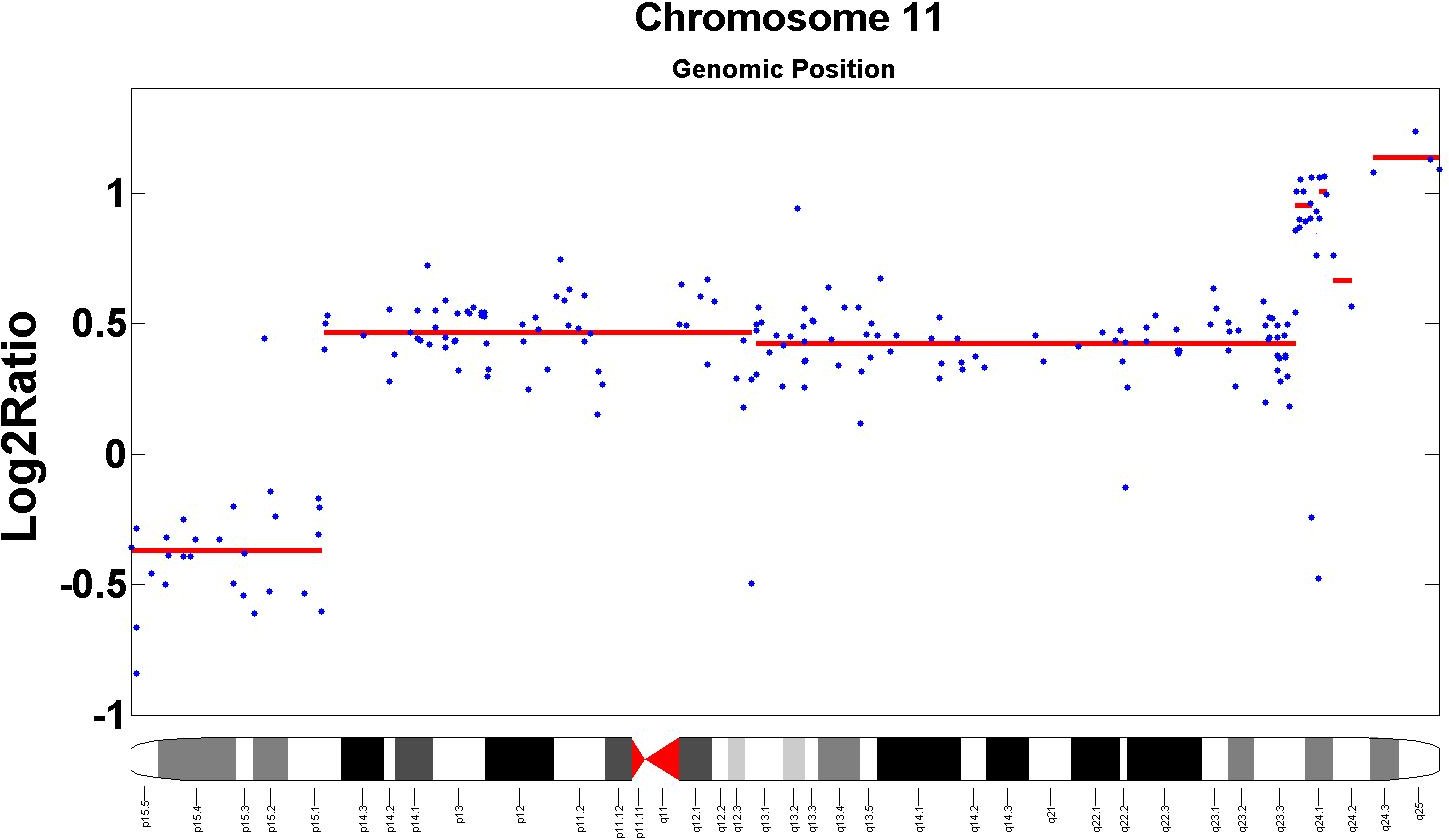} & \includegraphics[width=0.33\textwidth]{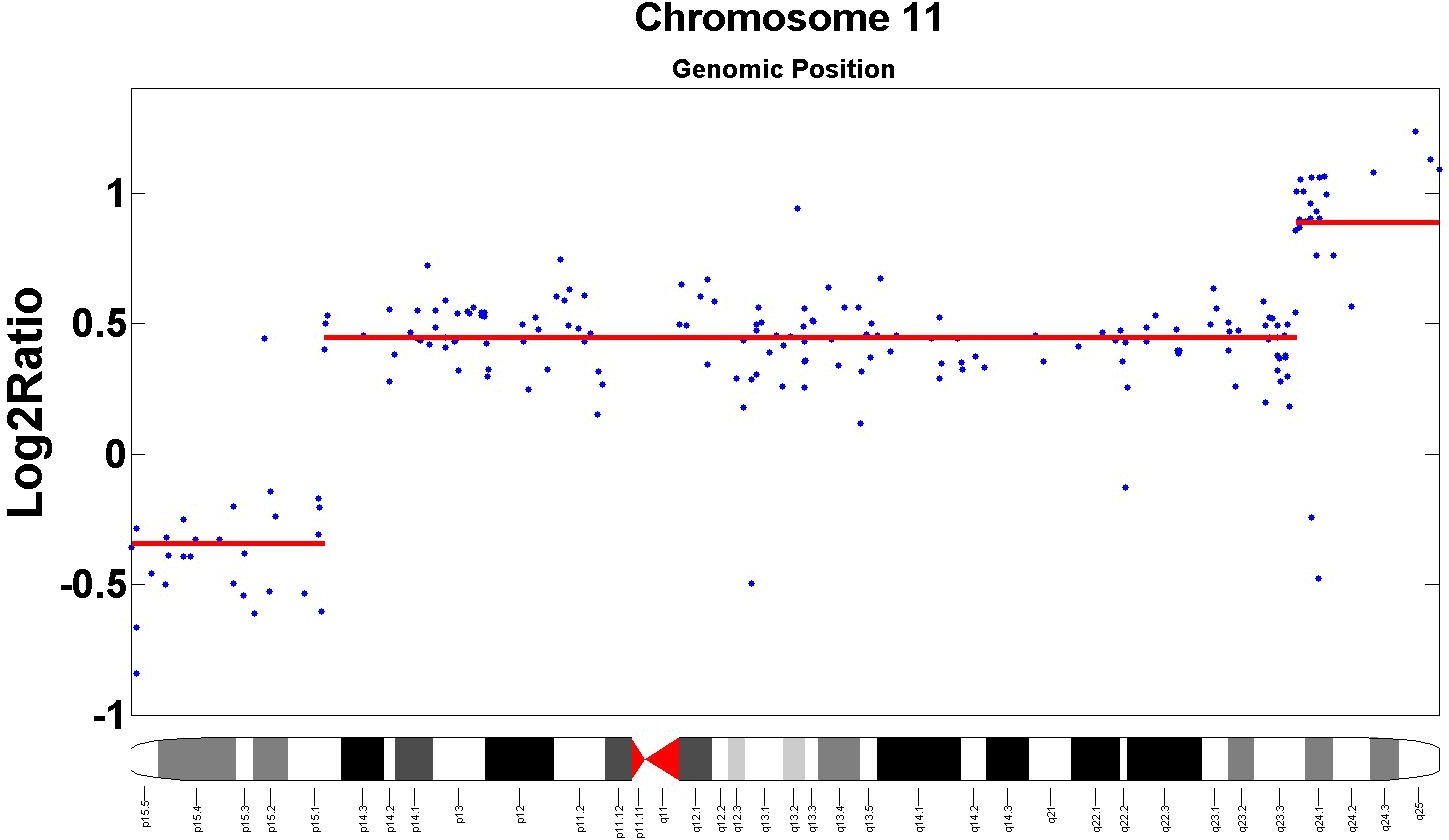} &    \includegraphics[width=0.33\textwidth]{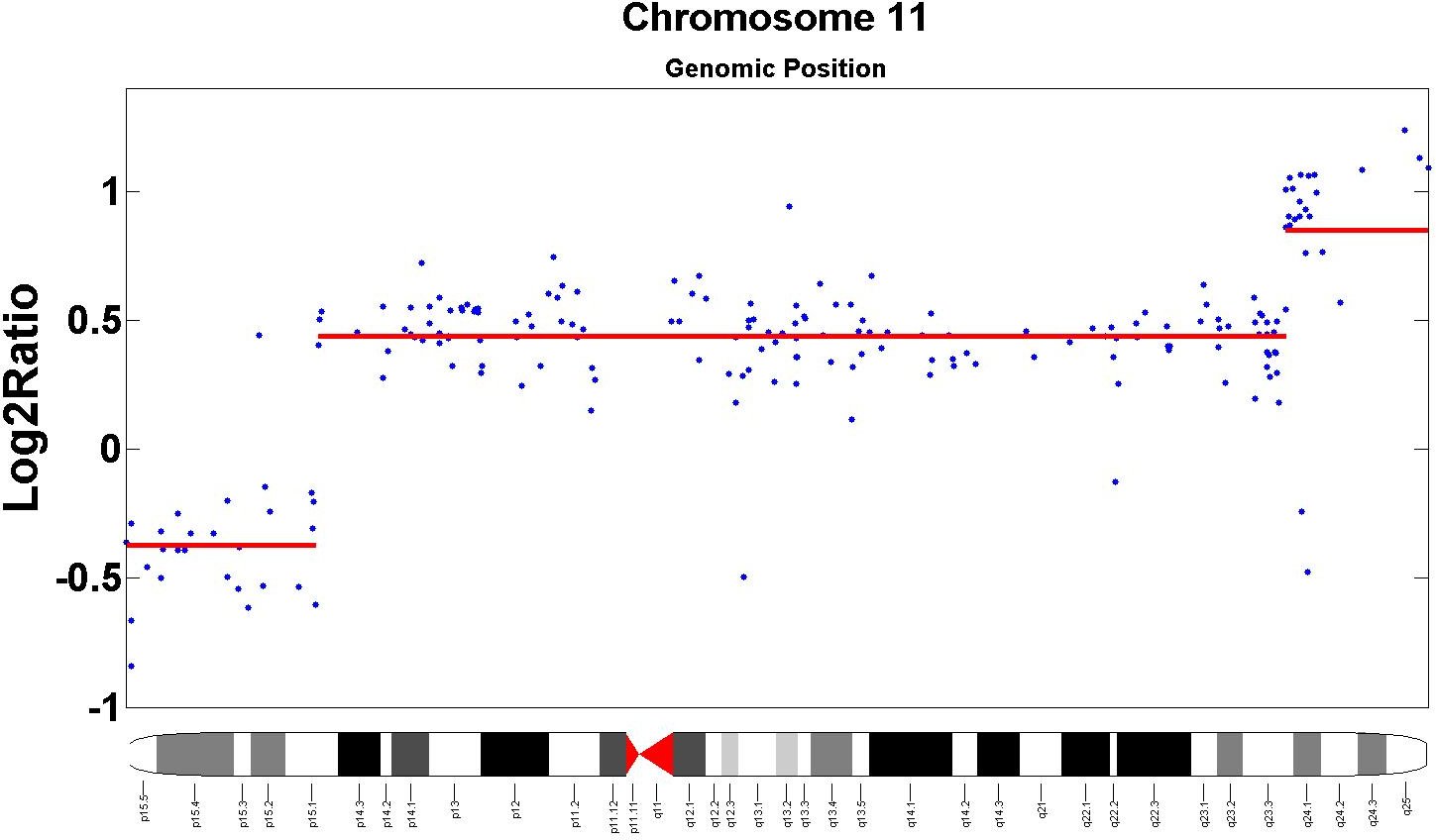} \\
		(d) &(e) &(f) \\
		\includegraphics[width=0.33\textwidth]{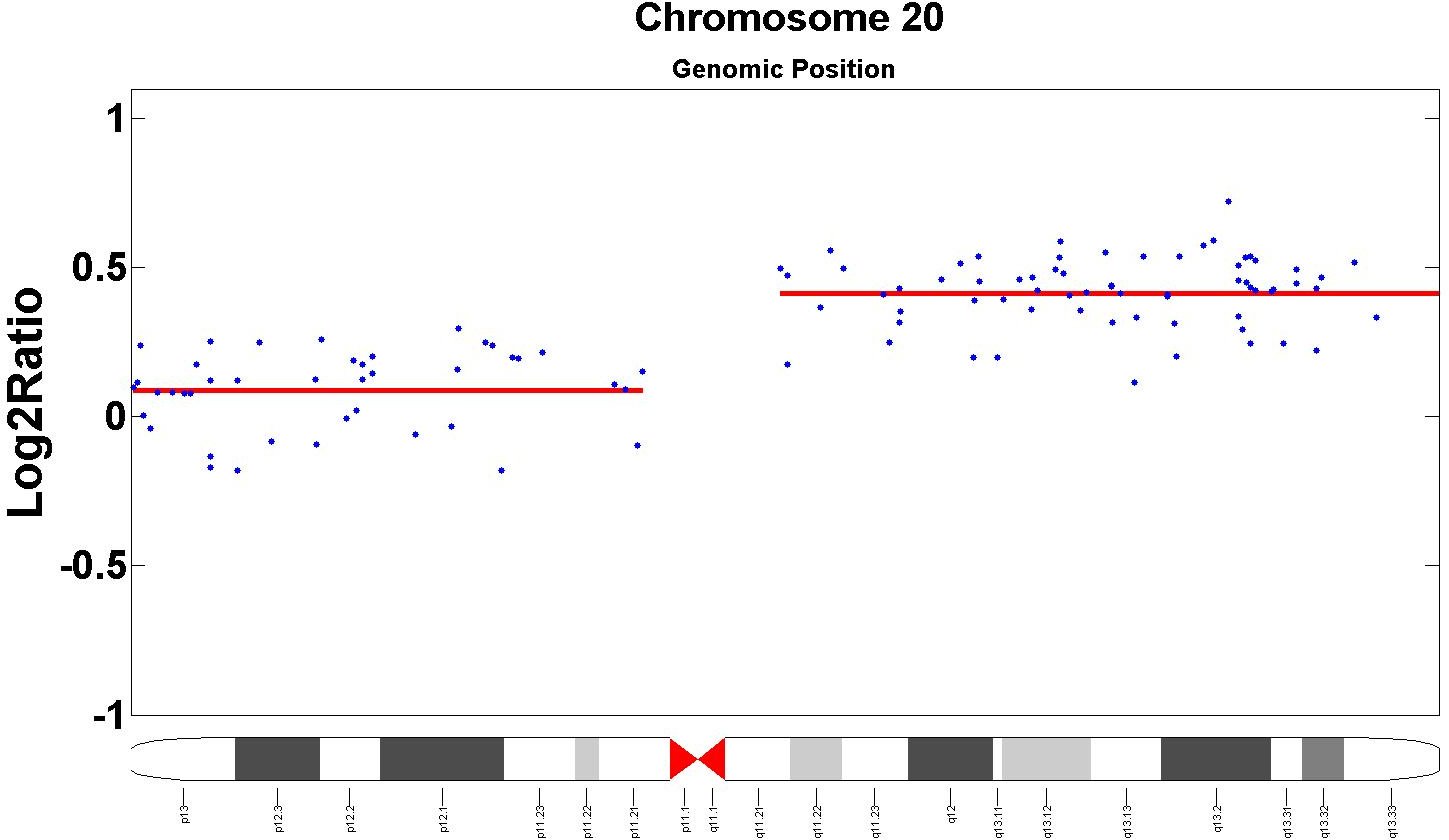} & \includegraphics[width=0.33\textwidth]{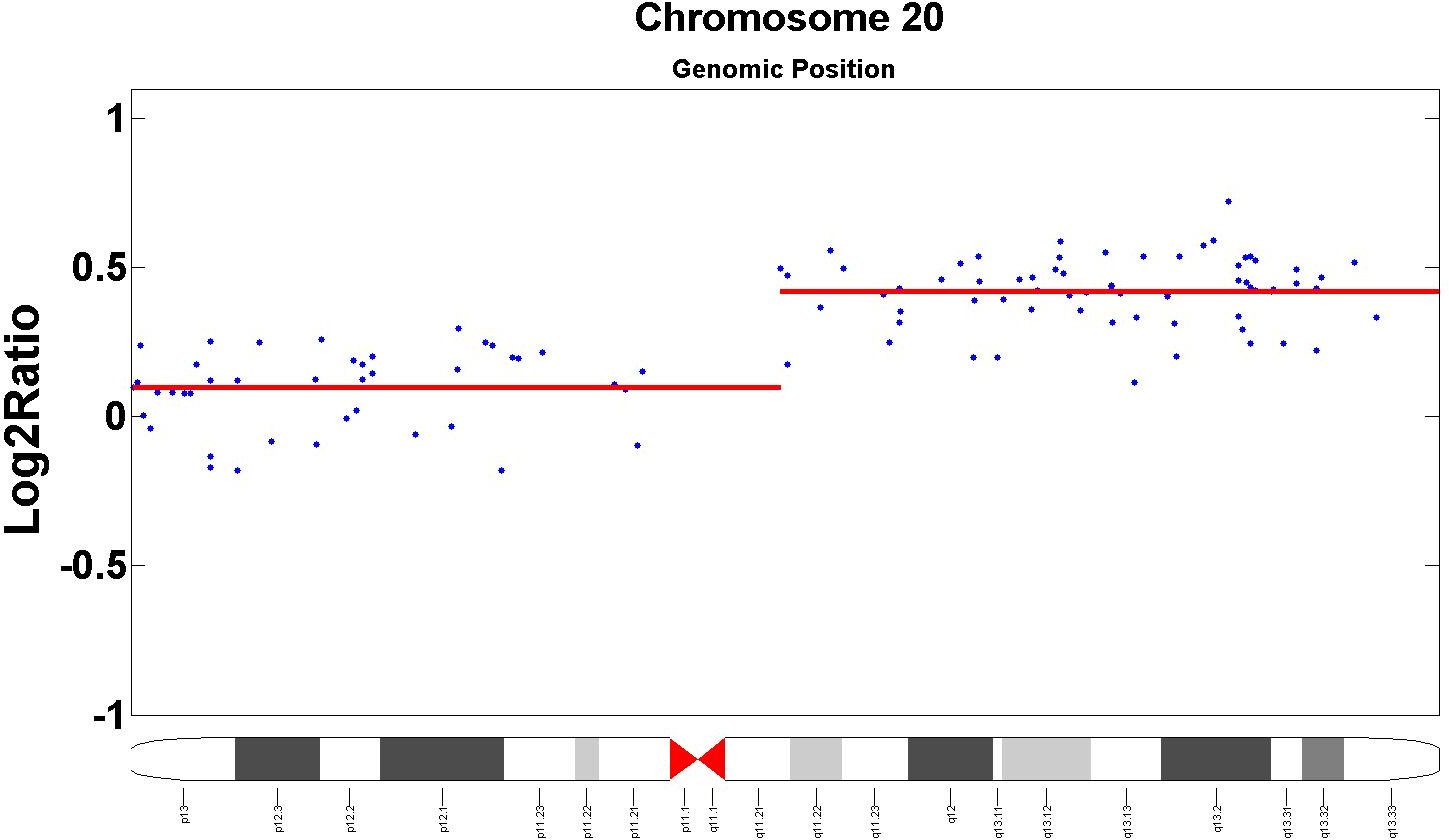} &    \includegraphics[width=0.32\textwidth]{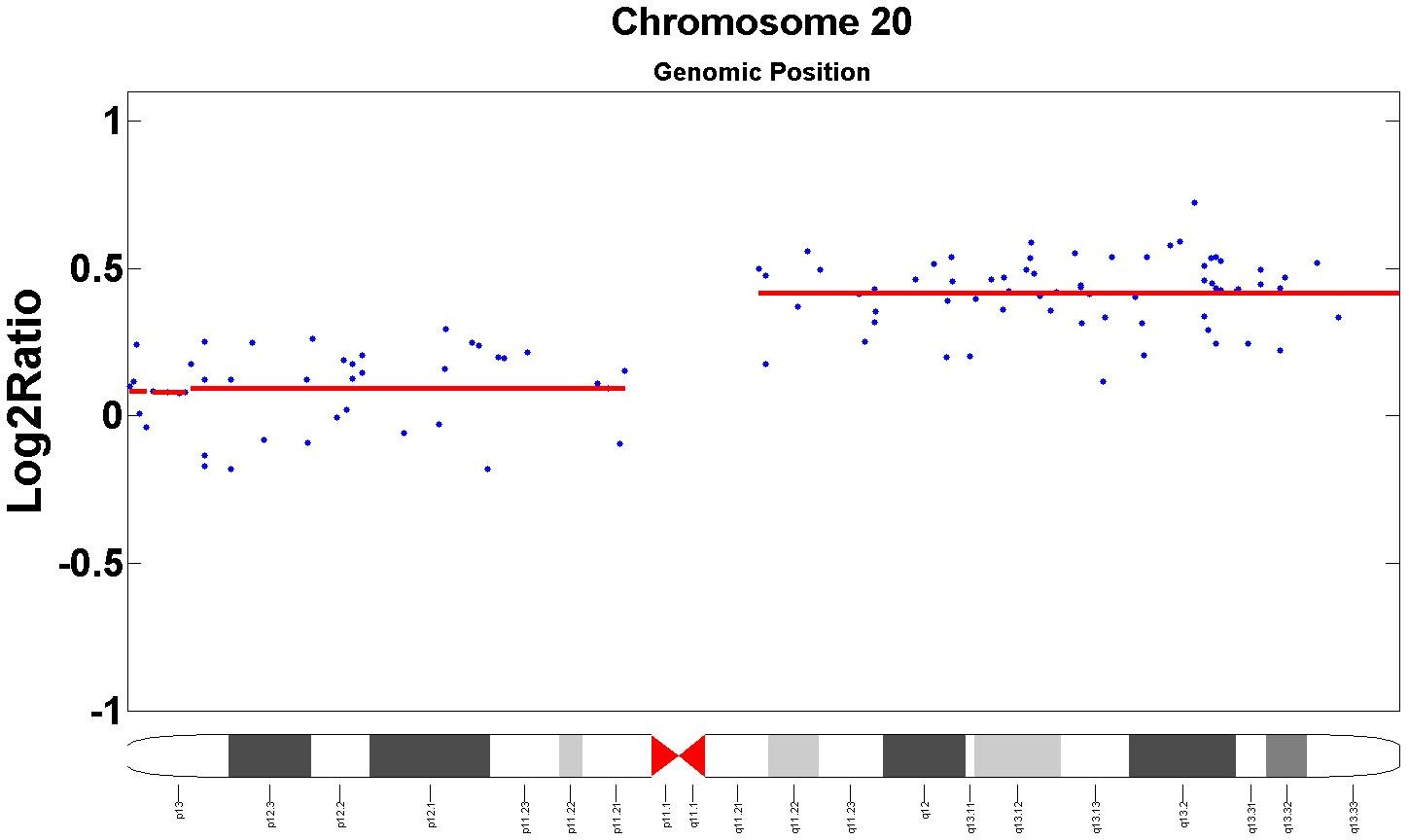} \\
		(g) &(h) &(i) \\
		\end{tabular}
\caption{Visualization of segmentation output of \trimmer, \dnacopy, and \cghseg\ for cell line T47D on chromosomes 1 (a,b,c), 11 (d,e,f), and 20 (g,h,i).  (a,d,g) \trimmer output.  (b,e,h) \dnacopy output. (c,f,i) \cghseg output.}
\label{fig:T47D}
\end{figure*}

On chromosome 17  (Figures~\ref{fig:HS578T}(g,h,i)), the three methods behave similarly, with all three predicting amplification of the p-arm.  \dnacopy\ places one more marker in the amplified region causing it to cross the centromere while \cghseg\ breaks the amplified region into three segments by predicting additional amplification at a single marker.

\paragraph{\textbf{Cell Line T47D:}}

Figure~\ref{fig:T47D} compares the methods on chromosomes 1, 8, and 20
of cell line T47D.  On chromosome 1 (Figure~\ref{fig:T47D}(a,b,c)),
all three methods detect loss of the p-arm and predominant
amplification of the q-arm.  \dnacopy\ infers a presumably spurious extension of the p-arm loss across
the centromere into the q-arm, while the other methods do not.  The main differences
between the three methods appear on the q-arm of chromosome 1.
\trimmer\ and \cghseg\ both detect a small region of gain proximal to
the centromere at 1q21.1-1q21.2, followed by a short region of loss
spanning 1q21.3-1q22. \dnacopy\ merges these into a single longer
region of normal copy number.  The existence of a small region of loss
at this location in breast cancers is supported by prior
literature~\cite{chunder}.

The three methods provide comparable segmentations of chromosome 11
(Figure~\ref{fig:T47D}(d,e,f)).  All predict loss near the p-terminus,
a long segment of amplification stretching across much of the p- and
q-arms, and additional amplification near the q-terminus.  \trimmer,
however, breaks this q-terminal amplification into several
sub-segments at different levels of amplification while \dnacopy\ and
\cghseg\ both fit a single segment to that region.  We have no
empirical basis for determining which segmentation is correct here.
\trimmer\ does appear to provide a spurious break in the long amplified
segment that is not predicted by the others.

Finally, along chromosome 20 (Figure~\ref{fig:T47D}(g,h,i)), the
output of the methods is similar, with all three methods suggest that
the q-arm has an aberrant copy number, an observation consistent with
prior studies \cite{breast20}.  The only exception is again that
\dnacopy\ fits one point more than the other two methods along the
first segment, causing a likely spurious extension of the p-arm's
normal copy number into the q-arm.

\section{Conclusions}
\label{sec:conc}
We have presented \trimmer, a new algorithm for detecting genomic regions of loss or gain in aCGH data. 
We compared \trimmer\ to two widely used methods, \dnacopy\ \cite{olshen} and \cghseg\ 
\cite{picard} that have previously been identified as the most successful among many options in the literature \cite{1181383}. \trimmer\ shows performance identical to \cghseg\ and superior to \dnacopy\ on a synthetic benchmark and superior performance to both on a benchmark of real cell line data.  Further demonstration of the methods on selected regions from a large breast cancer cell line dataset suggests that \trimmer\ is generally
superior at detecting fine-scale variations in aCGH data, while avoiding apparently spurious or misplaced breakpoints assigned by the other methods.  Where results differ between the methods, there is usually good support in the literature for the \trimmer\ segmentation.
Furthermore, \trimmer\ achieves superior accuracy with run times more than 50-fold faster than \cghseg\ and more than 500-fold faster than \dnacopy.

\section*{Acknowledgement} 

The authors would like to thank  F.~Picard for providing the MATLAB for the method of \cite{picard} and Ayshwarya Subramanian for helpful discussions.  

\paragraph{Funding} The authors were supported in this work by U.S. National Institutes of Health award \# 1R01CA140214.

\bibliographystyle{abbrv}
\bibliography{BIB/cghref}

\end{document}